\documentclass[a4paper, 10pt,oneside]{article}
\usepackage[T1]{fontenc}
\usepackage[english]{babel}
\usepackage{amsmath}
\usepackage{booktabs}
\usepackage{amsfonts}
\usepackage{mathtools}
\usepackage{tikz-cd}
\usepackage{sidecap}
\usepackage{amsmath}
\usepackage{mathrsfs}
\usepackage[font=scriptsize]{caption}
\usepackage{tabularx,ragged2e}
\usepackage{mwe} 
\usepackage{braket}
\usepackage{geometry}
\usepackage{subcaption}
\usepackage{yfonts}
\usepackage[toc,page]{appendix}
\usepackage{graphicx} 
\usepackage{wrapfig}
\usepackage{algorithm}
\usepackage{caption}
\usepackage{cite}
\usepackage{hyperref}
\hypersetup{colorlinks=false}
\usepackage{algpseudocode}
\usepackage{fancyhdr}

\begin{document}
  \date{\today}
  \thispagestyle{empty}
\begin{center}
Doc. n. P1800365\\
$\quad$\\
$\quad$\\
$\quad$\\
\Large
\textbf{IMR Consistency Tests with Higher Modes \\ on Gravitational Signals from the Second \\ Observing Run of LIGO and Virgo}\\
$\quad$\\
$\quad$\\
$\quad$\\
\large
Matteo Breschi$^{1,2,3}$, Richard O'Shaughnessy$^{4}$, Jacob Lange$^{4}$\\
and Ofek Birnholtz$^{4}$\\
$\quad$\\
$\quad$\\
$\quad$\\
\small
$^{1}$\emph{Dipartimento di Fisica "Enrico Fermi", Universit\`a di Pisa,}\\
\emph{Pisa I-56127, Italy}\\
$^{2}$\emph{Istituto Nazionale di Fisica Nucleare, Sezione di Pisa,}\\
\emph{Pisa I-56127, Italy}\\
$^{3}$\emph{Theoretisch-Physikalisches Institut, Friedrich-Schiller-Universit\"at,}\\
\emph{Jena, D-07743, Germany}\\
$^{4}$\emph{Center for Computational Relativity and Gravitation, Rochester Institute of Technology,}\\
\emph{Rochester, New York 14623, U.S.A.}\\
$\quad$\\
$\quad$\\
{March 14, 2019}
$\quad$\\
$\quad$\\
 \end{center}
 {\centering
\textbf{Abstract}\\}
Current tests of General Relativity are performed using approximations which neglect a key feature of complete solution of Einstein's theory: higher-order modes. Our analysis will reassess these tests, including these higher-order mode effects. We have chosen to perform inspiral-merger-ringdown consistency tests on the gravitational transients detected by LIGO and Virgo during the observing run O2. We use an approximant which includes all higher modes with $\ell \le 4$ (\texttt{NRSur7dq2}) and then, for the most interesting cases, we repeat the tests involving fits on Numerical Relativity simulations.

\newpage

\normalsize
 
\tableofcontents
\listoffigures
\newpage

\section{Introduction}

\par{These last few years have decreed the dawn of gravitational waves (GW) astronomy thanks to the combined work of LIGO Scientific and Virgo Collaboration (LVC), verifying the predictions of Einstein's theory of General Relativity (GR).}

\par{The first gravitational signal, GW150914 \cite{gw150914}, was detected on September 14th, 2015 and corresponds to a coalescence of binary black holes (BBH) with masses of $36^{+5}_{-4}\,M_\odot$ and $29^{+4}_{-4}\,M_\odot$ located at a luminosity distance of $410^{+160}_{-180}\,$Mpc. After this event, many other compact binary coalescences (CBC) of BBH have been detected \cite{gw151226 ,gw170104,  gw170608, gw170814} and on August 2017 the laser interferometers observed also a binary neutron stars (BNS) merger \cite{gw170817}. In the further years the sensitivity of the instruments will increase \cite{aLIGO, adVirgo} and numerous observations are expected \cite{cbcrate , cbcrate2}; it follows that a solid knowledge of the theoretical models and a robust method of parameter estimation are necessary in order to be able to understand the physics behind these events.}

\par{Generally, for data analysis purposes, the CBC signals are divided into three stages. Initially the two objects are rotating one around the other (\emph{inspiral}): the orbital frequency slowly increases while the distance between the two objects decreases because the system is losing energy due to the emission of GWs. Subsequently they approach and merge together (\emph{merger}): during this phase the field has large curvature, the orbital frequency increases sharply and we found the energy peak of the gravitational radiation. After the merger, a single remnant object is generated and it dissipates the residual energy through perturbative motions (\emph{ringdown}), emitting GWs though quasi-normal oscillating modes.}

\par{For a BBH coalescence, we need a set of 15 parameters to identify a single signal and we can divide those in two sub-sets: the \emph{intrinsic} parameters and the \emph{extrinsic} parameters. The intrinsic parameters are the fundamental to the description of the binary, if we change any intrinsic parameters we must recompute the orbital dynamics. We will call them as $\text{\boldmath$\xi$}$ and they are the masses $m_1$, $m_2$ (or any combination of these two variables) and the spins $\mathbf{S}_1$, $\mathbf{S}_2$. Usually the dimensionless spins parameters {\boldmath$\chi_1$}, {\boldmath$\chi_2$} are defined for a binary coalescence; they are such that $\text{\boldmath$\chi$}_{i}=c\mathbf{S}_i/Gm_i^2$ for $ \,i =1,2$. On the other hand, the extrinsic parameters simply describe how the binary is oriented in space and time relative to the detector; changing extrinsic parameters involves an easy transformation (rotation, translation or rescaling). For the further ones we will use the notation $\text{\boldmath$\lambda$}$  and they are the time at which the peak of the wave arrives at the Earth's geocenter $t_0$, the orbital phase of the binary at coalescence $\phi_0$, the right ascension $\alpha$ and the declination $\delta$ of the source, the inclination angle between the line of sight and the binary's angular momentum $\iota$, the luminosity distance $D$ and the polarization angle $\psi$. So, the complete set of parameters is $\text{\boldmath$\theta$}=\{\,$\text{\boldmath$\xi$}$\,,\,$\text{\boldmath$\lambda$}$\,\}$. }

\par{The observations in Ref.~\cite{gw150914, gw151226 ,gw170104,  gw170608, gw170814, gw170817} lead the scientific community to perform several test of GR, since thanks to GWs we are able to verify directly the predictions of Einstein's theory. Quite a few tests of GR currently performed on the BBH events detected by the ground-based interferometers of LIGO and Virgo can be found in Ref.~\cite{o2_tgr}. However, these tests are performed using approximations which neglect higher-order modes, which are an important amount of information of the complete solutions of gravitational radiation.}

\par{In this analysis, we present the results of IMR consistency tests on the O2 events including all higher modes up to $\ell \le 4$. We use \texttt{RapidPE} code \cite{rapidpe_pankow, rapidpe_lange} to evaluate the posterior distributions of BBH parameters. Initially we involve the numerical surrogate approximant in Ref.~\cite{surr_blackman} labelled \texttt{NRSur7dq2}. Then, for the most interesting cases, we repeat the tests using Numerical Relativity (NR) simulations from the RIT catalog in Ref.~\cite{rit_catalog}. After reviewing the theoretical models and the data analysis involved methods, we describe the results of IMR tests on the strains observed by LIGO and Virgo detectors during the observation run O2, which measured the gravitational waves produced by compact binary coalescences. These events are labelled as GW170104, GW170809, GW170814, GW170823 and GW170818. The others O2 events, that we did not study (GW170608 and GW170817), have low masses and thus SNR is too low in the two portions of signal (inspiral and post-inspiral), preventing the IMR consistency test from being performed. We also do not include GW170729 in this work.}

\par{Let us introduce some basic notions about parameter estimation. The GWs' parameters estimation is based on the comparison between a template $h(t,\text{\boldmath$\theta$})$ (which depends on some parameters $\text{\boldmath$\theta$}$) and the observed strain $s(t)$. If we can suppose that a signal is present within a data segment, we are able to decompose into the inherent noise of the detector and a gravitational-wave signal,
\begin{align}
\label{match_filt}
s(t)=h(t,\text{\boldmath$\theta$}) + n(t)\,,
\end{align}
where $n(t)$ denotes the noise inside the detector. Under these assumptions, the raw strain of data is transformed in the Fourier's space and each frequency is weighted on the  power spectral density (PSD). The PSD is a characteristic function of the single interferometer and it describes the intensity of the noise in the detector at a given frequency. Under the assumption of Gaussian and stationary noise, we can define the (one-sided) PSD $S_n(f)$ as
\begin{align}
\langle \tilde n(f)\, \tilde n(f')\rangle=\frac{1}{2}\,S_n(f)\,\delta(f-f')\,.
\end{align} 
We note that this average should be performed over many possible realizations of the system; however we have only a single physical system (i.e. our detector) but we can follow it in time, so the ensemble average is replaced by the time average. Another way to extract the PSD pass from the auto-correlation function $R(\tau)$,
\begin{align}
R(\tau)=\langle n(t+\tau)\, n(t)\rangle=\int_0^\infty S_n(f)\,e^{2\pi i f\tau}\,df\,.
\end{align}
From these results, it follows that the PSD is the quantity that measures our sensitivity. Then, assuming a gaussian stationary zero-mean noise, we compare the data with the template, i.e. we minimize the detector-noise-weighted residuals expressed as a conventional gaussian (log) likelihood for the data in the presence of a signal. The noise weighting is expressed using an inner product,
\begin{align}
\label{inner_product}
(a|b)=\int _{0}^{\infty} \frac{ \tilde a^*(f)\,\tilde b(f) + \tilde a(f)\,\tilde b^*(f)}{S_n(f)}\,df
\end{align}
with $a = b = s-h$. This definition is a natural consequence of the auto-correlation $R(\tau)$ inside the detector with the assumption of zero-mean stationary noise. In order to compare signals coming from different detectors, the waveform is shifted in reference to a detector relative to the geocentric time. Furthermore, this alignment between signals from different detectors gives us informations regarding the sky location of the source. Once we minimize the noise-weighted residuals, we are able to infer what kind of signal was detected and we get the posterior distributions of those parameters. From Eq.~\eqref{inner_product} the definition of signal-to-noise ratio (SNR) follows,
\begin{align}
\text{SNR}[h]=\frac{(s|h)}{\sqrt{(h|h)}}
\end{align}
which quantify the quality of our detected signal weighted on the noise inside the detector. For GW150914 it was reached an SNR approximately equal to 24, which is an incredible large value.}\

\section{Higher Modes}

\par{In general, we can decompose the gravitational strain in oscillation modes using the 2-spin-weighted spherical harmonics ${}^{(-2)}{Y}_{\ell ,m }$,
\begin{align}
\label{spher_harm_decompos}
h=h_+ + ih_\times =\frac{1}{D}  \sum_{\ell\ge 2}\sum_{m=-\ell}^{\ell} {}^{(-2)}{Y}_{\ell, m }(\iota,\phi_0)h_{\ell ,m}(\text{\boldmath$\xi$},t)\,,
\end{align}
where $h_{\ell m}$ denotes the $(\ell,m)$ mode of the wave. In this section, we will focus on binaries in a quasi-circular orbit in the $z=0$ plane (e.g., without precession); in this formalism we can write,
\begin{align}
h_{\ell, m}(\text{\boldmath$\xi$},t) = (-1)^\ell h^*_{\ell, m}(\text{\boldmath$\xi$},t)\,,
\end{align}
and decomposing the single mode into real amplitude and real phase $h_{\ell m}=A_{\ell m}\cdot e^{i \phi_{\ell m}}$, we get
\begin{align}
\phi_{\ell, m}(t) = m \,\phi_\text{orb}(t)\,,
\end{align}
where $\phi_\text{orb}$ denote the orbital phase of the binary.}

\par{The spin-weighted spherical harmonics $^{(s)}Y_{\ell,m}(\vartheta,\varphi)$ (where $\vartheta$ and $\varphi$ are used as generic angles) are defined in terms of the Wigner functions, as metioned in Ref.~\cite{pnmodes_kidder},
\begin{align}
^{(s)}Y_{\ell,m}(\vartheta,\varphi) = (-1)^{-s}\sqrt{\frac{2\ell+1}{4\pi}} \,d_{ms}^\ell(\vartheta)\,e^{im\,\varphi},
\end{align}
where
\begin{align}
\begin{split}
d_{ms}^\ell(\vartheta) = &\sqrt{(\ell+m)!(\ell-m)!(\ell+s)!(\ell-s)!}\\
&\times\sum_{k=k_0}^{k_f}\frac{(-1)^k \big(\sin\frac{\vartheta}{2}\big)^{2k+s-m}\big(\cos\frac{\vartheta}{2}\big)^{2\ell+m-s-2k}}{k!(l+m-k)!(l-s-k)!(s-m+k)!}\,,\\
\end{split}
\end{align}
where $k_0=\max{(0,m-s)}$ and $k_f=\min{(\ell+m,\ell-s)}$.
}

\par{Furthermore, in Ref.~\cite{multipole_thorne}, another analogue decomposition is used, involving the mass-type ${U}^{\ell,m}$ and the current-type ${V}^{\ell,m}$ multipole moments,
\begin{align}
\label{multipole_decompos}
h_{ij}^\text{TT}=\frac{G}{c^2D}\mathbb{P}_{ij,mn}\sum_{\ell=2}^\infty \sum_{m=-\ell}^\ell \Bigg[\frac{1}{c^\ell}{U}^{\ell,m}(t_\text{ret})\,T_{ij}^{\text{E2}, \ell,m}+\frac{1}{c^{\ell+1}}{V}^{\ell,m}(t_\text{ret})\,T_{ij}^{\text{B2}, \ell,m}\Bigg]\,.
\end{align}
The tensors $T_{ij}^{\text{E2}, \ell,m}$ and $T_{ij}^{\text{B2}, \ell,m}$ are pure-spin tensor harmonics, and can be derived from the 2-spin weighted spherical harmonics by
\begin{align}
\label{pure-spin-tensor}
T_{ij}^{\text{E2}, \ell,m}=\frac{1}{\sqrt{2}}\bigg( {}^{(s)}Y_{\ell,m}e_ie_j + {}^{(-s)}Y_{\ell,m}e_i^* e_j^* \bigg)\,,\\
T_{ij}^{\text{B2}, \ell,m}=\frac{-i}{\sqrt{2}}\bigg( {}^{(s)}Y_{\ell,m}e_ie_j - {}^{(-s)}Y_{\ell,m}e_i^* e_j^* \bigg)\,,
\end{align} 
where $e_i=(u_i+iv_i)/\sqrt{2}$ and $u_i,v_i$ are previously defined. Combining Eq.~\eqref{spher_harm_decompos},~\eqref{multipole_decompos} and~\eqref{pure-spin-tensor}, we get
\begin{align}
h^{\ell,m}=\frac{G}{\sqrt{2}\, c^{\ell+2}}\bigg[U^{\ell,m}(t_\text{ret})-\frac{i}{c}V^{\ell,m}(t_\text{ret})\bigg]
\end{align}}

\par{As we can see in Ref.~\cite{hm_bustillo}, the higher modes (HMs) contributions are relevant in the last few orbits of the large mass ratio $q=m_1/m_2$ ($q\ge1$), i.e. when one object is more massive than the other. In Ref.~ \cite{hm_bustillo}, the effects due to precession are highlighted: precessing simulations show a triggered amplitude, and the same modulations are present in the waves' frequency evolution. 

\par{The BBH we are going to study (GW170104, GW170809, GW170814, GW170818 and GW170823) are characterized by a low mass ratio $q<4$ and mass $M=m_1+m_1<100M_\odot$, as we can see in Ref.~\cite{gwtc1}. For such sources the $(2,\pm2 )$ modes dominate the sum of all components. Moreover the efficiency of sub-dominant modes is strongly correlated with the inclination angle $\iota$, as we can see in Eq.~\eqref{spher_harm_decompos}. However the sub-dominant modes' contributions increase with the angle $\iota$.}

\par{In general, the dominant mode gave us a sufficient description of the events. Our purpose is to verify these statements, comparing the results of previous GR tests, made up without the HM contributions, with our analysis, which include the HM effects. In order to do that we will perform an IMR consistency test on the O2 events.}

\subsection{NR Surrogate: \texttt{NRSur7dq2}}

\par{For our purposes, we need an approximant which includes HM at least up to $\ell = 4$. A good approximant is given by J. Blackman \emph{et al.} in Ref.~\cite{surr_blackman}. In this articles, it is presented an NR surrogate, called \texttt{NRSur7dq2}, which is able to describe the wave evolution for BBH with $q<2$ and for $|\mathbf{\chi}_{1,2}|<0.8$ . This surrogate also include the precessing contributes, however, for the purposes of this report, we will only use the surrogate to describe non-precessing binaries }

\par{This surrogate model is made up using parameters fitted on several NR simulations. Specifically, since GWs are highly oscillatory and they change in complicated ways as one varies masses and spins and since we have to interpolate the model in a high-dimensional space, the template is decomposed in \emph{waveform data pieces}. Each waveform data piece is a simpler function that varies slowly over parameters. Once each waveform data piece is interpolated over a desired set of points in parameter space, $h(t)$ is recombined from these pieces. Actually, in order to conserve the continuity and the differentiability of the physical quantities, the templates are built up using a set of differential equations, computing them into different frames (co-precessing and co-orbital). This equations are those that describe the evolution of spins, frequencies and phase of the binary.}

\par{As mentioned in Ref.~\cite{surr2_blackman, surr3_blackman}, the $m=0$ modes of \texttt{NRSur7dq2} do not attempt to reproduce the expected memory terms. However, memory modes are very low frequency and therefore are not a significant contribution to the results of the parameter estimations in this search.}

\section{Parameters Estimation}

\par{Let us introduce some of the fundamental concepts for Bayesian inference. Calling $s$ the data strain, $h$ the model (based on certain hypothesis $\mathcal{H}$) and {\boldmath{$\theta$}} the parameters' vector, the Bayes' theorem states that
\begin{equation}
\label{bayes_theo}
p(\text{\boldmath{$\theta$}}|s, \mathcal{H})=\frac{p(s|\text{\boldmath{$\theta$}}, \mathcal{H})p(\text{\boldmath{$\theta$}}|\mathcal{H})}{p(s| \mathcal{H})}\,,
\end{equation}
where $p(\text{\boldmath{$\theta$}}|s, \mathcal{H})=\mathcal{P}(\text{\boldmath{$\theta$}})$ is the posterior distribution of the parameters given an observation, $p(s|\text{\boldmath{$\theta$}}, \mathcal{H})=\mathcal{L}(\text{\boldmath{$\theta$}})$ is the likelihood function of the data, $p(\text{\boldmath{$\theta$}}|\mathcal{H})=\Pi(\text{\boldmath{$\theta$}})$ is the prior distribution of the parameters assumed before the measurements and $p(s| \mathcal{H})=\mathcal{Z}$ is the normalization constant called evidence. }

\par{Let us call $\mathcal{H}_0$ the hypothesis for which there is no signal inside the observed strain and call $\mathcal{H}_1$ the one that assume a gravitational signal inside the strain. So, assuming \eqref{match_filt} and supposing that we have a set of $k$ detectors, we get
\begin{align}
\label{lnL_signal}
\mathcal{H}_0:\,s_k=n_k\quad&\Rightarrow\quad\ln\mathcal{L}(\mathcal{H}_0)\propto -\sum_k \big(s\big|s\big)_k\,,\\
\mathcal{H}_1:\,s_k=h_k-n_k\quad&\Rightarrow\quad\ln\mathcal{L}(\text{\boldmath{$\theta$}}|\mathcal{H}_1)\propto -\sum_k \big(s-h(\text{\boldmath{$\theta$}})\big|s-h(\text{\boldmath{$\theta$}})\big)_k\,,
\end{align}
where we used the inner product defined in Eq.~\eqref{inner_product}. Then, thanks to the Bayes' theorem \eqref{bayes_theo}, we can compute the evidences $\mathcal{Z}_0$ and $\mathcal{Z}_1$, and we are able to infer that a gravitational is contained in the strain $s$ if $\mathcal{Z}_1>\mathcal{Z}_0$. We start from the assumption that our analyzed strain contain a GW, as verified by the previous LVC analysis. }

\par{We note that the output of the detector is a time series which describes the oscillations of the test-masses, while a GW is a tensorial perturbation $h_{ij}(t)$. However, the tensorial signal is reduced to a scalar due to the detector; in fact we can writhe for the $k$-th detector
\begin{align}
\label{antenna_pattern}
h_k(t,\text{\boldmath$\theta$})&=F_{k,ij}\,h_{k,ij}(t,\text{\boldmath$\theta$})\\
\notag&=F_{k,+}(\alpha,\delta,\psi)\, h_{k,+}(t,\text{\boldmath$\theta$})+F_{k,\times} (\alpha,\delta,\psi)\, h_{k,\times}(t,\text{\boldmath$\theta$})\,,
\end{align} 
where $F_{+,\times}$ are the antenna pattern functions, that describe the sensitivity of the instruments in the different directions, and $h_{+,\times}$ are the two polarization of the gravitational strain, which depends also on the position of the $k$-th detector with respect to the geocenter. Then we define $F_k=F_{+,k}+iF_{\times,k}$ and combining Eq.~\eqref{antenna_pattern} and Eq.~\eqref{spher_harm_decompos}, we get
\begin{equation}
\label{rapidpe_decomp}
h_k(\text{\boldmath{$\theta$}},t) = \Re\,\frac{F_k(\alpha,\delta,\psi)}{D}\sum_{\ell, m }h_{\ell,m}(\text{\boldmath{$\theta$}},t) {}^{(2)}Y_{\ell,m}(\iota,\phi_0)\,.
\end{equation}
For a set of $k$ detectors, we can write the time at which the peak of the GW arrives in the $k$-th detector as
\begin{align}
t_k=t_0+\frac{\mathbf{x}_k\cdot \mathbf{n}(\alpha,\delta)}{c}\,,
\end{align}
where $\mathbf{x}_k$ is the vector from the geocenter to the $k$-th detector and $\mathbf{n}$ is the unitary wave vector that depends on the sky position of the source. Thanks to the comparison of the different time delays we are able to infer on the sky location of the source, and it emerge that only using three detectors we are able to localize the source in a sufficient narrow spot. In fact, using only two detectors, we are not able to fix an angular degree of freedom, and so the posterior distribution is spread over a circumference.}

\subsection{\texttt{RapidPE}}

\par{In order to perform these tests we use the parameters estimation software \texttt{RapidPE}, described in Ref.~\cite{rapidpe_pankow, rapidpe_lange}. This code is based on Monte Carlo processes that perform fast computation of the likelihood over an input grid, and process these results returning a posterior output file. Then, the output file could be used as new grid for a second iteration (over the same observation), in order to get an output posterior distribution that is localized in a narrower region. }

\par{The efficiency of \texttt{RapidPE} is due to how it treats the different sets of parameters. We recall that the entire set of parameters $\text{\boldmath{$\theta$}}$ can be divided into two subsets of extrinsic $\text{\boldmath{$\lambda$}}$ and intrinsic $\text{\boldmath{$\xi$}}$ parameters. The waveform decomposition in Eq.~\eqref{rapidpe_decomp} can be exploited to speed up evaluations of likelihood. In fact, the decomposed modes $h_{\ell,m}$ depends only on the intrinsic parameters $\text{\boldmath{$\xi$}}$ and the extrinsic ones (except for $t_k$) are encoded in the coefficients of the linear combination. So, we can take out from the inner product these parameters, thus we need only to compute the inner product involving $h_{\ell,m}$ and the data strain $s$. So, we define the quantities 
\begin{align}
\mathcal{Q}_{k,\ell,m}(\text{\boldmath{$\xi$}},t_k)=\big( h_{\ell,m}(\text{\boldmath{$\xi$}},t_k)\big| s \big)_k=2 \int_0^\infty df\,\frac{\tilde h^*_{\ell,m}(\text{\boldmath{$\xi$}},f)\,\tilde s(f) }{S_{n,k}(f)}\,e^{2\pi i f t_k} \,,
\end{align}
\begin{align}
\mathcal{U}_{k,\ell,m,\ell',m'}(\text{\boldmath{$\xi$}})=\big(h_{\ell,m}\big|h_{\ell',m'}\big)_k\,,
\end{align}
\begin{align}
\mathcal{V}_{k,\ell,m,\ell',m'}(\text{\boldmath{$\xi$}})=\big(h^*_{\ell,m}\big|h_{\ell',m'}\big)_k\,.
\end{align}
A signal will produce a peak in the filtered outputs localized to a short millisecond time window around the coalescence time, so $\mathcal{Q}_{k,\ell,m}$ will be sharply peaked as functions of $t_k$ and we need only retain the values for a narrow range of $t_k$. To allow for detector arrival times that differ from the geocenter time, the range of $t_k$ for which we must store the $\mathcal{Q}_{k,\ell,m}$  is set by the light travel time across Earth ($2R_\oplus/c\approx 42 $ms).}

\par{Now, plugging together Eq.~\eqref{rapidpe_decomp} and the definition of likelihood, we have
\begin{align}
\begin{split}
\ln\mathcal{L}(\text{\boldmath{$\xi$}},\text{\boldmath{$\lambda$}})\propto&\frac{1}{D}\sum_k\sum_{\ell,m}\big(F_k{}^{(-2)}Y_{\ell,m}\big)^* \mathcal{Q}_{k,\ell,m}(\text{\boldmath{$\xi$}},t_k)\\
&-\frac{1}{4D^2}\sum_k\sum_{\ell,m,\ell',m'}\bigg[ |F_k|^2\, {}^{(-2)}Y_{\ell,m}^* {}^{(-2)}Y_{\ell',m'}\,\mathcal{U}_{k,\ell,m,\ell',m'}(\text{\boldmath{$\xi$}})\\
&\qquad\qquad\qquad\qquad+F_k^2 \,{}^{(-2)}Y_{\ell,m} {}^{(-2)}Y_{\ell',m'}\,\mathcal{V}_{k,\ell,m,\ell',m'}(\text{\boldmath{$\xi$}}) \bigg]\,.
\end{split}
\end{align}
Importantly, the intrinsic parameters $\text{\boldmath{$\xi$}}$ enter only through the $\mathcal{Q}_{k,\ell,m}$, $\mathcal{U}_{k,\ell,m,\ell',m'}$ and $\mathcal{V}_{k,\ell,m,\ell',m'}$. These are the dominant cost, as they require computing the orbital dynamics, the $h_{\ell,m}$, inner product integrals, and inverse Fourier transforms. By contrast, the extrinsic parameters enter the $F_k$ and ${}^{(-2)}Y_{\ell,m}$, which are much cheaper to compute.}

\par{Technically, \texttt{RapidPE} is composed by two steps: the first integrate out the extrinsic parameters and we call it \emph{integrate-likelihood-extrinsic} (\texttt{ILE}), the second step processes the \texttt{ILE}'s output to generate the posterior samples and it is called \emph{compute-intrinsic-posterior} (\texttt{CIP}). In the following sections we will give an idea of these processes.}

\subsubsection{\texttt{ILE}}

\par{The first step integrate the extrinsic parameters $\text{\boldmath{$\lambda$}}$, because precomputed quantities allow us to an efficiently evaluation of $\mathcal{L}(\text{\boldmath{$\xi$}},\text{\boldmath{$\lambda$}})$ as a function of $\text{\boldmath{$\xi$}}$ defined as
\begin{align}
\label{ile_integral}
\mathcal{L}(\text{\boldmath{$\xi$}})=\int\mathcal{L}(\text{\boldmath{$\xi$}},\text{\boldmath{$\lambda$}})\Pi(\text{\boldmath{$\lambda$}})d\text{\boldmath{$\lambda$}}\,,
\end{align}
where $\Pi(\text{\boldmath{$\lambda$}})$ is the prior distribution for the extrinsic parameters. We assume the sources analyzed are randomly oriented and randomly distributed in the Universe out to a fiducial radius. With the potential exception of the sky position, our priors are independent, and thus separable. This computation is performed over a grid inscribed into the prior's bounds and the algorithm uses a Monte Carlo iteration.}

\par{Moreover, in order to improve the efficiency, a weight probability function $w(\text{\boldmath{$\lambda$}})$ is used in the extrinsic parameters and it must be different from zero in the entire domain. Then the integral Eq.~\eqref{ile_integral} is rewrite as
\begin{align}
\label{ile_integral_w}
\mathcal{L}(\text{\boldmath{$\xi$}})=\int\frac{\mathcal{L}(\text{\boldmath{$\xi$}},\text{\boldmath{$\lambda$}})\Pi(\text{\boldmath{$\lambda$}})}{w(\text{\boldmath{$\lambda$}})}\,\Big[w(\text{\boldmath{$\lambda$}})d\text{\boldmath{$\lambda$}}\Big]\,,
\end{align}
and this functions is used also to compute the marginalized distributions for the extrinsic parameters (necessary to integrate them out). The \texttt{ILE} output file is a list of events in the intrinsic parameters' space which describe the $\ln\mathcal{L}$ function. Then, this points are post-processed and collected together and the output is given to the \texttt{CIP}.}

\subsubsection{\texttt{CIP}}

\par{Once we have evaluated the the likelihood over a grid of extrinsic parameters, $\mathcal{L}(\text{\boldmath{$\xi$}})$ is computed via Gaussian process interpolation, over the intrinsic parameters space. This step generates a grid with a density that conforms to the values of the intrinsic likelihood., i.e. more points where the likelihood's values are larger. Then, the algorithm compute the evidence,
\begin{align}
\mathcal{Z}=\int \mathcal{L}(\text{\boldmath{$\xi$}})\Pi(\text{\boldmath{$\xi$}})d\text{\boldmath{$\xi$}}\,,
\end{align}
and the posterior's values over the current grid,
\begin{align}
\mathcal{P}(\text{\boldmath{$\xi$}})=\frac{\mathcal{L}(\text{\boldmath{$\xi$}})\Pi(\text{\boldmath{$\xi$}})}{\mathcal{Z}}\,.
\end{align}
To extract the posterior samples, the software executes a fit on the data and then an adaptive Monte Carlo on these results getting an independent set of values from the posteriors. When fitting the likelihood, we employ coordinate systems well-adapted to the likelihood, which are likely to produce an approximately gaussian likelihood in the limit of strong signals. We set the number of output samples equal to 8000 points.}

\par{To construct posteriors for the intrinsic parameters, we adopt a uniform prior in masses and spins during the first iteration. For the next ones, we use the likelihood values obtained with previous runs as input grid, and then perform an analysis with the same priors. Actually, this is the real power of \texttt{RapidPE}; we can re-use the output of \texttt{CIP} as new input for the \texttt{ILE}, and re-process the same values, obtaining a peaked distribution over a sufficient narrow region of the parameters' space. When the correlations between the posterior samples does not change increasing the number of iterations, we are able to suppose that we reach the convergence of the distribution. In the end, we compare our results with the analysis presented in Ref.~\cite{o2_tgr}.}

\section{IMR Consistency Tests}

\par{The IMR consistency test perform independent parameter estimations on the inspiral portion of data and on the post-inspiral one. This test is based on estimating the mass and the spin of the remnant black hole from the two independent portions of signal. Then, the results, coming from the low-frequency inspiral and the high-frequency post-inspiral, are compared by checking that they are consistent, as it should be from GR predictions. This test was introduced by A. Ghosh \textit{et al.} in Ref.~\cite{tgr_bbh_ghosh, tgr_bbh_ghosh2} and Fig.~\ref{ghosh_plots} shows their results obtained using simulated events.}

\par{In order to perform the test, we have to choose the cut off frequency, that divides the inspiral from merger and ringdown, since we perform the analyses in the frequency domain. We define the inspiral [post-inspiral] as the Fourier frequencies lower [greater]  than that of the innermost stable circular orbit (ISCO) of a Kerr black hole with mass $M_f$ and spin $\chi_f$. However, this choice is not unique  and reasonable alternatives do not have a significant effect on the test, if SNR is sufficiently large in both stages. In order to be consistent with the results in Ref.~\cite{o2_tgr}, we will choose $f_\text{cut}$ equal to those chosen during the previous analyses. The only exception is GW170814, where we move $f_\text{cut}$ to 150 Hz instead of 161 Hz.}

\par{Once we choose the cutoff frequency, we select a model and we use \texttt{RapidPE} software, explained in Ref.~\cite{rapidpe_pankow,rapidpe_lange }, to perform parameter estimations on the two data segments (inspiral and post-inspiral) measuring the posterior distributions of the intrinsic parameters and we repeat the same for the entire signal (labelled as IMR). Then we involve the routine of the IMR tests~\cite{imr_location} looking for the consistency between the two portions of the signal. We expect to found agreement between these results, if GR theory's predictions are valid. As mentioned by Ref.~\cite{tgr_bbh_ghosh, tgr_bbh_ghosh2}, IMR tests are performed on the final observables values, i.e. mass and spin parameter of the remnant object. We will use to compute these quatitites the NR fit formulae given by Ref.~\cite{final_mass_spin_healt, final_spin_hbr, final_mass_spin_uib}. This function is located in \texttt{LAL}~\cite{lal_dir} and it is labelled as \texttt{bbh\_average\_fits\_precessing}.}

\begin{SCfigure}
\includegraphics[width=9cm]{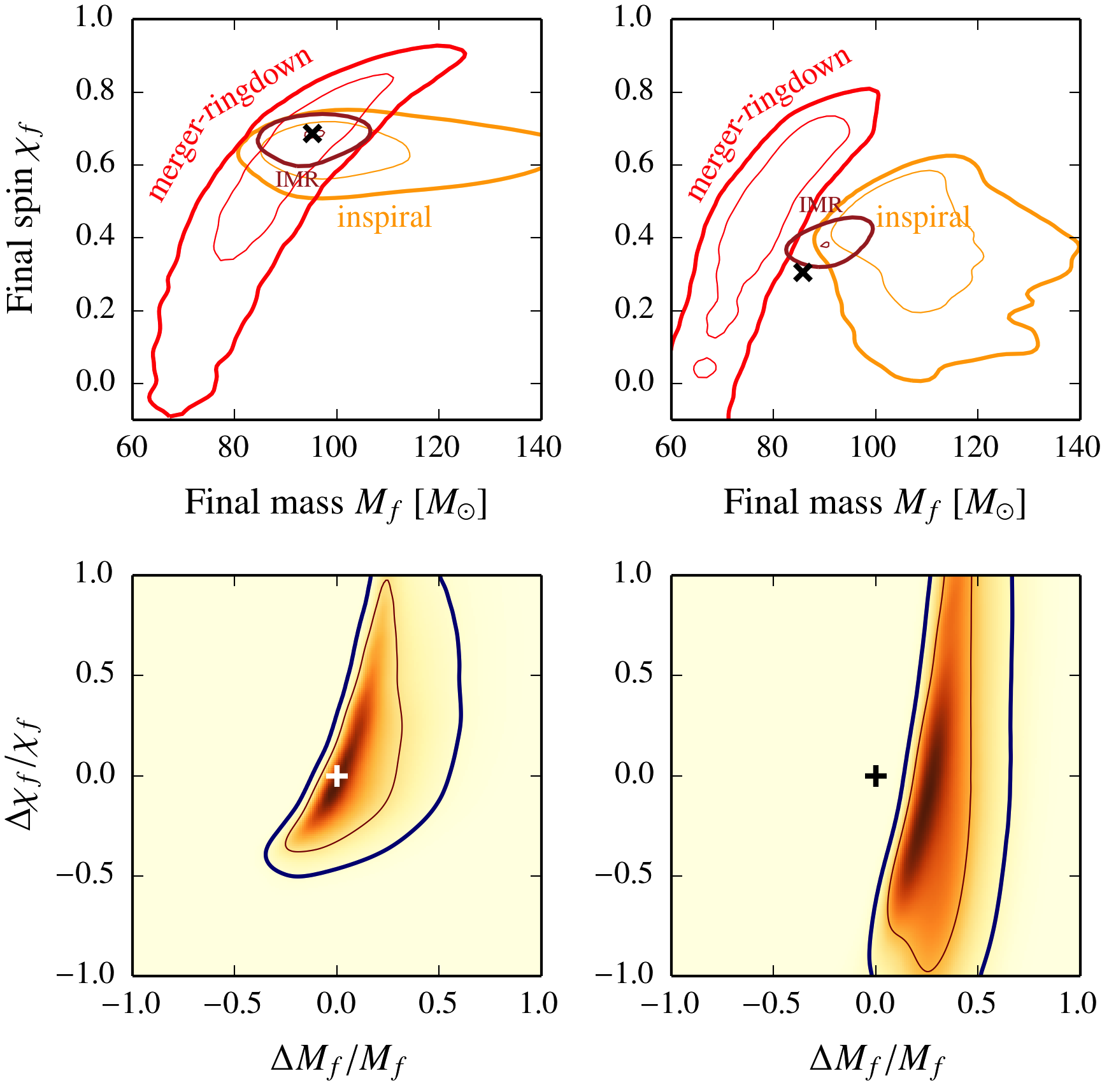}
\caption[IMR consistency test of simulated events from Ref.~\cite{tgr_bbh_ghosh}]{The figure is Fig.1 of A. Ghosh \textit{et al.}, Phys. Rev. D 94, 021101(R) (Ref.~\cite{tgr_bbh_ghosh}) and it shows the results of an IMR consistency test. This analysis is performed on simulated non-spinning GR event with $m_1=m_2 = 50\,M_\odot$ and optimal SNR of 25 in the advanced LIGO detectors. On the left, the top panel shows the 68\% and 95\% credible regions of the posterior distributions of the mass and spin of the final black hole estimated from the in-spiral and post-inspiral parts of a simulated GR signal, respectively; the bottom left panel shows the posterior of the parameters $\Delta M_f/\bar M_f$, $\Delta \chi_f /\bar \chi_f$t hat describe the deviation from GR, estimated from the same simulation. On the right, same as the left panels, except that here the injection corresponds to a kludge modified GR injection, as highlighted by the plots where the GR value is well outside the 95\% credible region. }
\label{ghosh_plots}
\end{SCfigure}

\begin{figure}[thb]
\centering
\includegraphics[width=7.3cm]{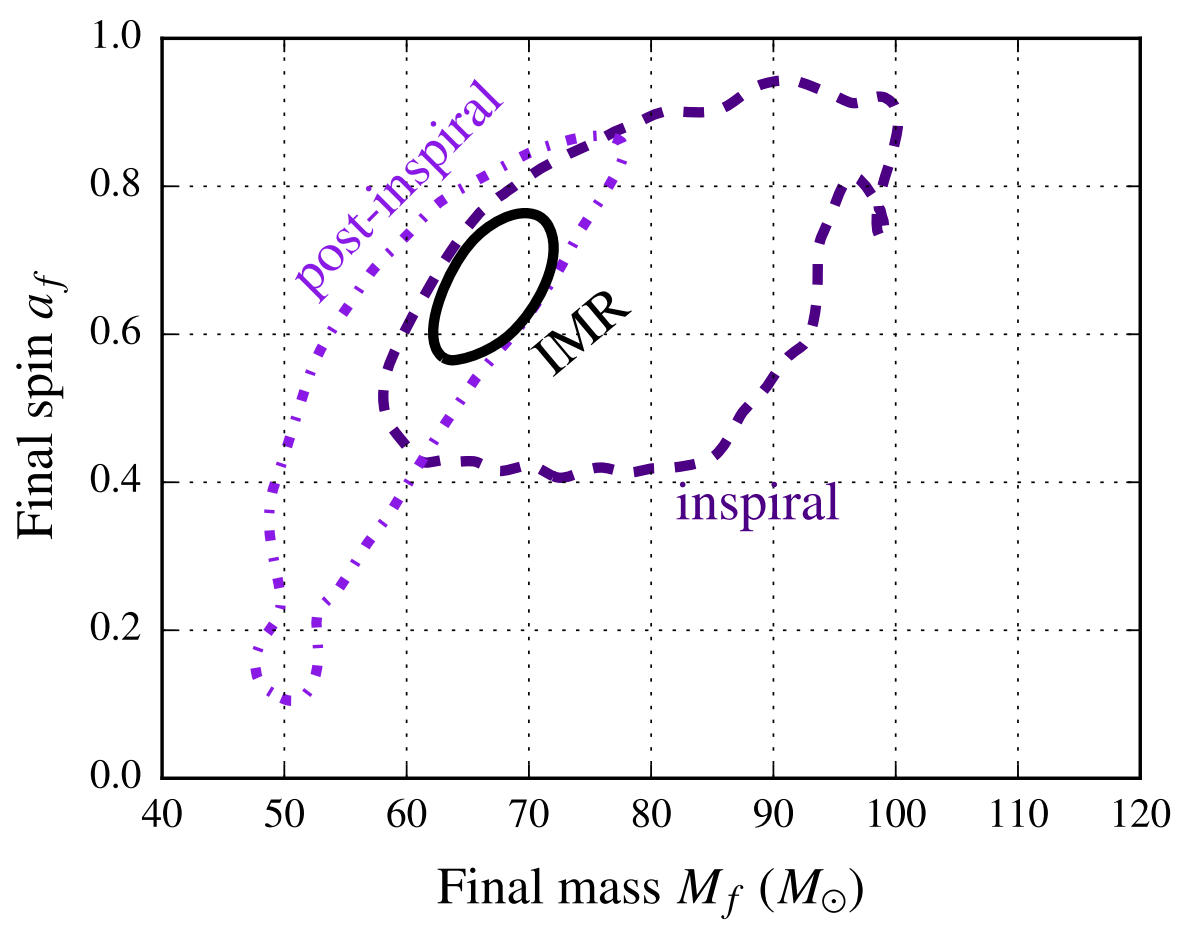}\includegraphics[width=7.3cm]{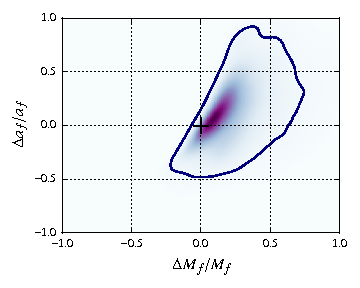}
\caption[IMR consistency test of GW150914 from Ref.~\cite{gw150914_tgr}]{The IMR test plots of GW150914 shown in Fig.4 of LIGO Scientific and Virgo Collaboration, Phys. Rev. Lett. 116, 221101 (Ref.~\cite{gw150914_tgr}). On the left, the 90\% credible regions of the joint probability distribution of final mass and final spin for the final object as determined from the inspiral, from the post-inspiral and from the entire IMR signals. On the right, the joint probability distributions of $\Delta M_f/\bar M_f,\, \Delta \chi_f\bar\chi_f$. The $+$ symbol indicates the null result expected in GR, which lies on the isoprobability contour that encloses 28\% of the posterior. In this analysis \texttt{IMRPhenomPv2}~\cite{imrphenom, imrphenom2, imrphenom3} and \texttt{SEOBNRv2\_ROM}~\cite{seobnrv4, seobnrv4rom} approximants are used to estimate the posterior distribution.}
\label{imrtgr_plots}
\end{figure}

\par{In Fig.\ref{imrtgr_plots} are shown the results of the IMR consistency test of GW150914 from LVC paper in Ref.~\cite{gw150914_tgr}; for the purpose of this test, the cutoff frequency is chosen equal to 132 Hz. On the left side, the inspiral posterior distribution is consistent with the post-inspiral one up to the 90\% confidence level, and in the common region we can found the IMR results. So we are able to infer that these results do not deviate from the prediction of BBH in GR. In order to assess the results the quantities $\Delta M_f/\bar M_f$ and $\Delta \chi_f/\bar \chi_f$ are defined. Those describe the fractional differences of the final masses and spins and they quantify the consistency of the observed signal with a BBH predicted by GR. In detail, these quantities are the differences between the inspiral and the post-inspiral measures divided by the averages of those, as explained in Ref.~\cite{tgr_bbh_ghosh2}. Explicitly,
\begin{align}
\Delta M_f = M_f^\text{I} - M_f^\text{MR}\quad , \quad \Delta \chi_f = \chi_f^\text{I} - \chi_f^\text{MR}\,,
\end{align}
and
\begin{align}
\bar M_f = \frac{M_f^\text{I} + M_f^\text{MR}}{2} \quad , \quad \bar \chi_f = \frac{\chi_f^\text{I} + \chi_f^\text{MR}}{2}\,
\end{align}
Where the labels I and MR denote respectively ``inspiral'' and ``merger-ringdown''. The GR's predictions coincide with the origin of the axes for these quantities, and this point is included in the posterior distribution on the isoprobability level that enclose 28\% of the posterior distribution.}

\par{We note that  we cannot perform this test on all O2 events because some of those have low masses and the SNR is too low for parameter estimation to provide sufficient information about the parameters from the two portion of signal (inspiral and post-inspiral). We can use only sufficiently massive BBH coalescence, because, with the current sensitivities of the instruments, we are not able to obtain acceptable posterior distributions with BNS and low-mass BBH mergers' data.}

\par{If we have a set of $N$ observation we can combine the posterior distributions fro the fractional quantities getting an overall posterior. Calling $\epsilon_M \equiv \Delta M_f /\bar M_f$ and $\epsilon_\chi \equiv \Delta \chi_f /\bar \chi_f$, it follows that for $N$ observations the combined posterior distribution for these fractional quantities is:
\begin{align}
\mathcal{P}\left(\epsilon_M,\epsilon_\chi \right) = \Pi\left(\epsilon_M,\epsilon_\chi \right) \prod_{i=1}^N \frac{\mathcal{P}_i\left(\epsilon_M,\epsilon_\chi\right)}{\Pi_i \left(\epsilon_M,\epsilon_\chi \right)}\,,
\end{align}
where $\Pi\left(\epsilon_M,\epsilon_\chi\right) $ is the overall prior distribution for the fractional quantities and $\Pi_i \left(\epsilon_M,\epsilon_\chi\right)$ are the priors used to compute the posterior $\mathcal{P}_i$. }

\subsection{IMR Tests on O2 Events}

\par{In the following sections, we show the IMR consistency tests performed on O2 events and using all HMs with $\ell\le 4$. The prior distributions for the extrinsic parameters are chosen flat over the entire domain, except for the distance $D$, which is proportional to $D^2$ in the range $[0, 3\, \text{Gpc}]$, the inclination angle $\iota$ which is proportional to $\sin\iota$ in its domain and the prior for the sky position $(\alpha, \delta)$ is isotropic over the solid angle $d\Omega = \sin\delta \, d\delta \,d\alpha$. Moreover we set the entire frequency range from 20 Hz to 1024 Hz, and we will split it into two separate ranges for the IMR test according to the estimated cut-off frequency. }

\par{Regarding the intrinsic parameters, we use an uniform prior for the masses $m_1$ and $m_1$, imposing the condition $q<2$ since NR surrogate model is not reliable over $q>2$. This choice is not a severe restriction since we do not expect to find events with large mass ratio, and this is also proved by the detections catalog GWTC-1 in Ref.~\cite{gwtc1}, where the events of interest have mass ratio sufficiently lower than 2. Under this point of view, GW170104 and GW170818 are an exceptions, since the 90\% isoprobability contour of the mass ratio posteriors
reach values respectively of 2.45 and 2.94 . However, for GW170104 we repeat the test involving pure NR simulations without any restriction on the mass ratio. Regarding the other details about the intrinsic parameters priors, we explain the chosen prior distributions in every sub-sections. The effects of the prior selection for Bayesian analysis affect the results, as shown by Ref.~\cite{prior_vitale}, and we have to use the correct distributions for each specific hypothesis. In order to simplify the analysis, the spins are taken aligned orthogonally to the orbit ($\chi_{i,x}=\chi_{i,y}=0$ for $ i=1,2$), and precession effects are neglected.}

\par{For the most interesting cases (GW170104, GW170814 and GW170823) we evaluate the posterior distributions involving fits on pure NR simulations. In this cases we do not have any restriction on masses and spins, but we still keep aligned components of the spins. These simulations are reliable from the frequency of 35 Hz, so we have to use a larger lower-frequency in order to avoid numerical errors and maintain the sanity of the waveforms. When we use pure NR fits we move the lower-frequency from 20 Hz to 35 Hz.}

\subsubsection{GW170104}

\par{We use a flat prior for the masses components taking $M$ in the range between $40\,M_\odot$ and $80\,M_\odot$ and the spins components' prior are in agreement with the aligned spins assumption. Our model \texttt{NRSur7dq2} includes all HMs up to $\ell=4$ and the chosen cut-off frequency is 143 Hz.}

\par{Fig~\ref{imrtgr_0104_surr} shows the join posterior samples for mass and spin parameter of the final black hole for the results with \texttt{NRSur7dq2} and the relative posterior for the fractional quantities $\Delta M_f/\bar M_f,\, \Delta \chi_f/\bar \chi_f$.  GR prediction is enclosed in the 46.3\% confidence level. In Fig.~\ref{imrtgr_0104_nr}, the same result with NR simulations fits; GR predictions lies on the isoprobability contour at the 1.9\%confidence level. 

\begin{figure}[tb]
\centering
\includegraphics[width=7.5cm]{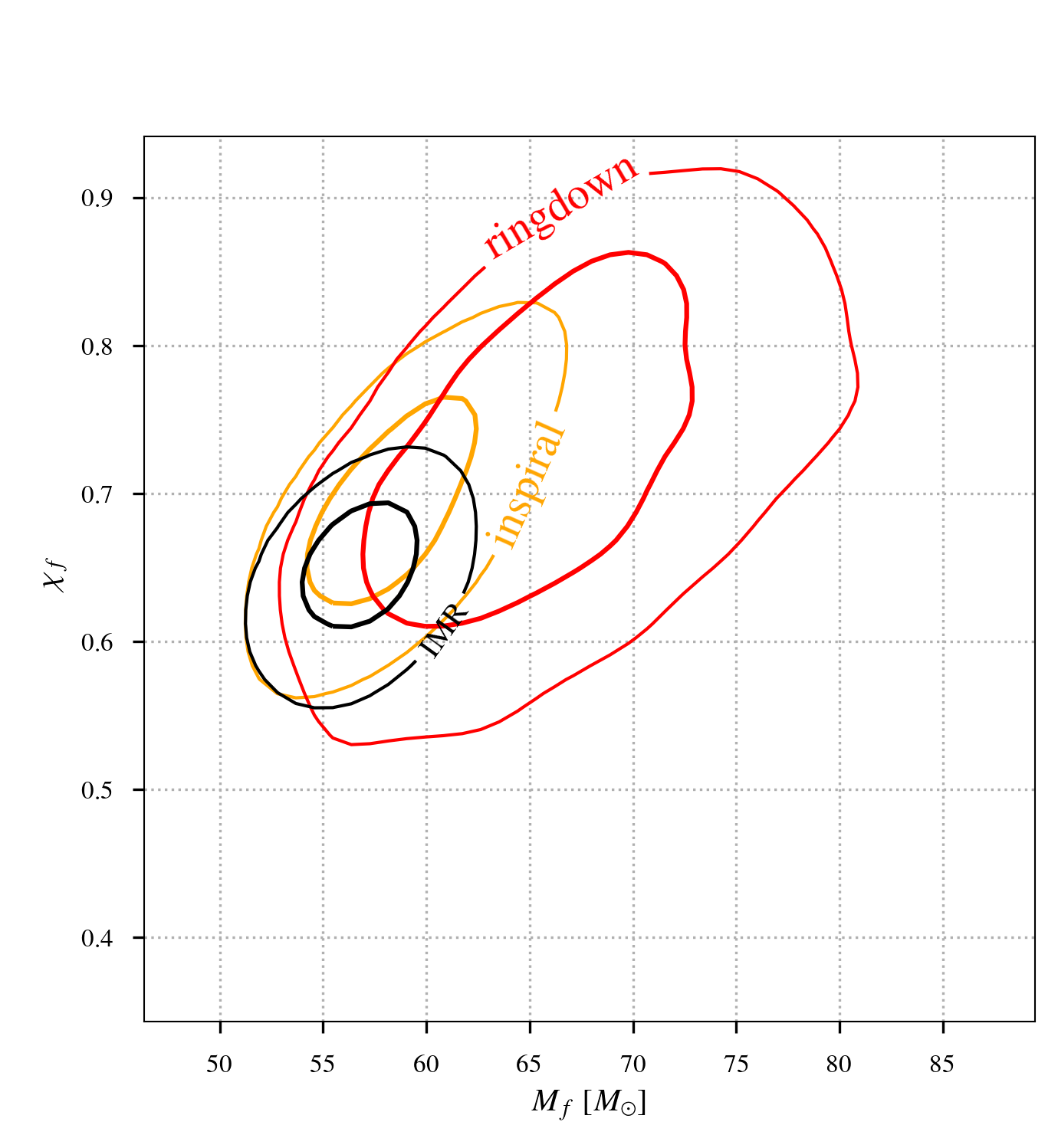}\includegraphics[width=7.5cm]{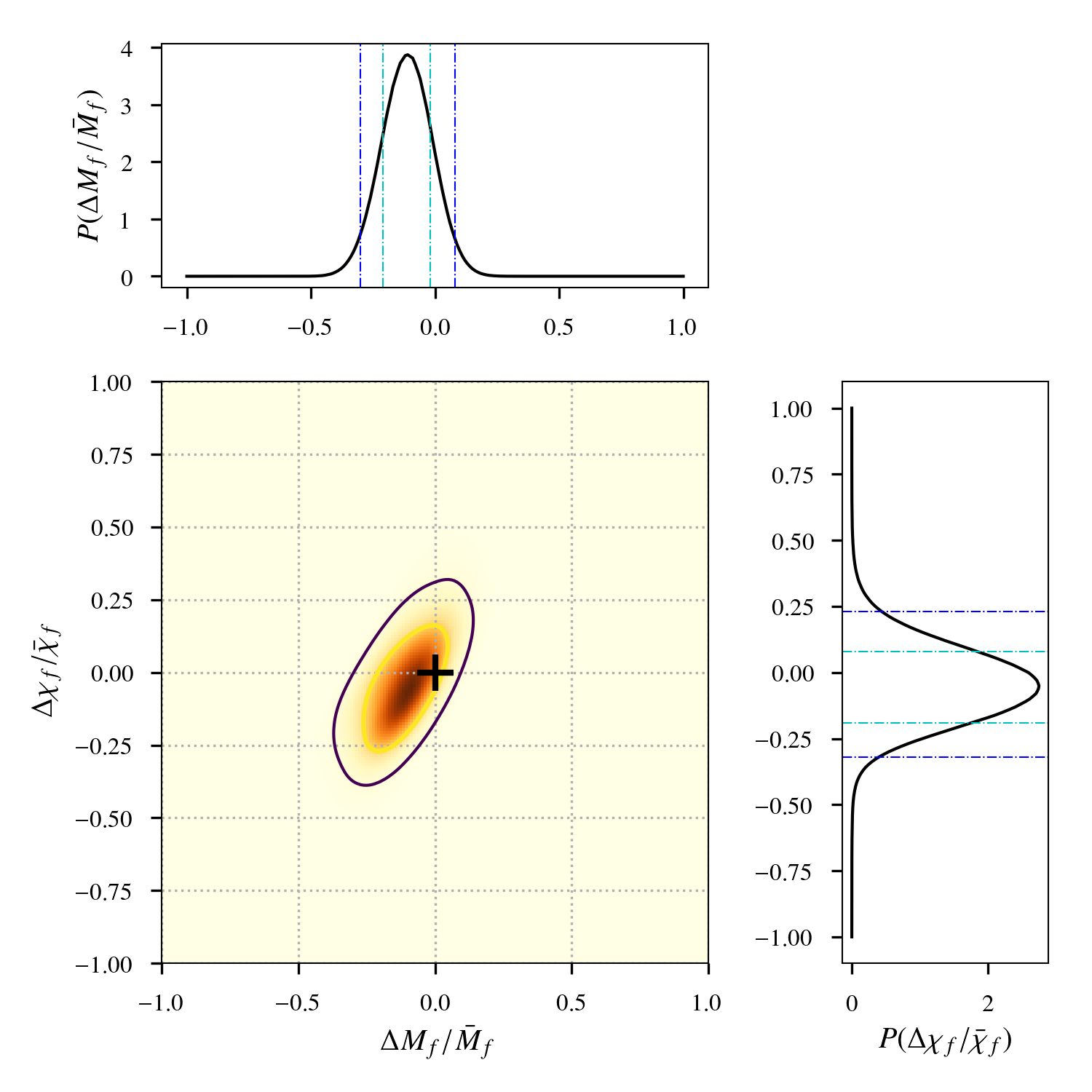}
\caption[IMR consistency test of GW170104 (NR surrogate)]{On the left, IMR overlap plot of GW170104 using all HMs up to $\ell = 4 $ with \texttt{NRSur7dq2} approximant assuming aligned spins; on the right, the joint probability distributions of $\Delta M_f/\bar M_f,\, \Delta \chi_f/\bar \chi_f$ does not show deviation from GR above 46.3\%. All the contours indicate the 68\% and the 95\% credible regions.}
\label{imrtgr_0104_surr}
\end{figure}

\begin{figure}[H]
\centering
\includegraphics[width=7.5cm]{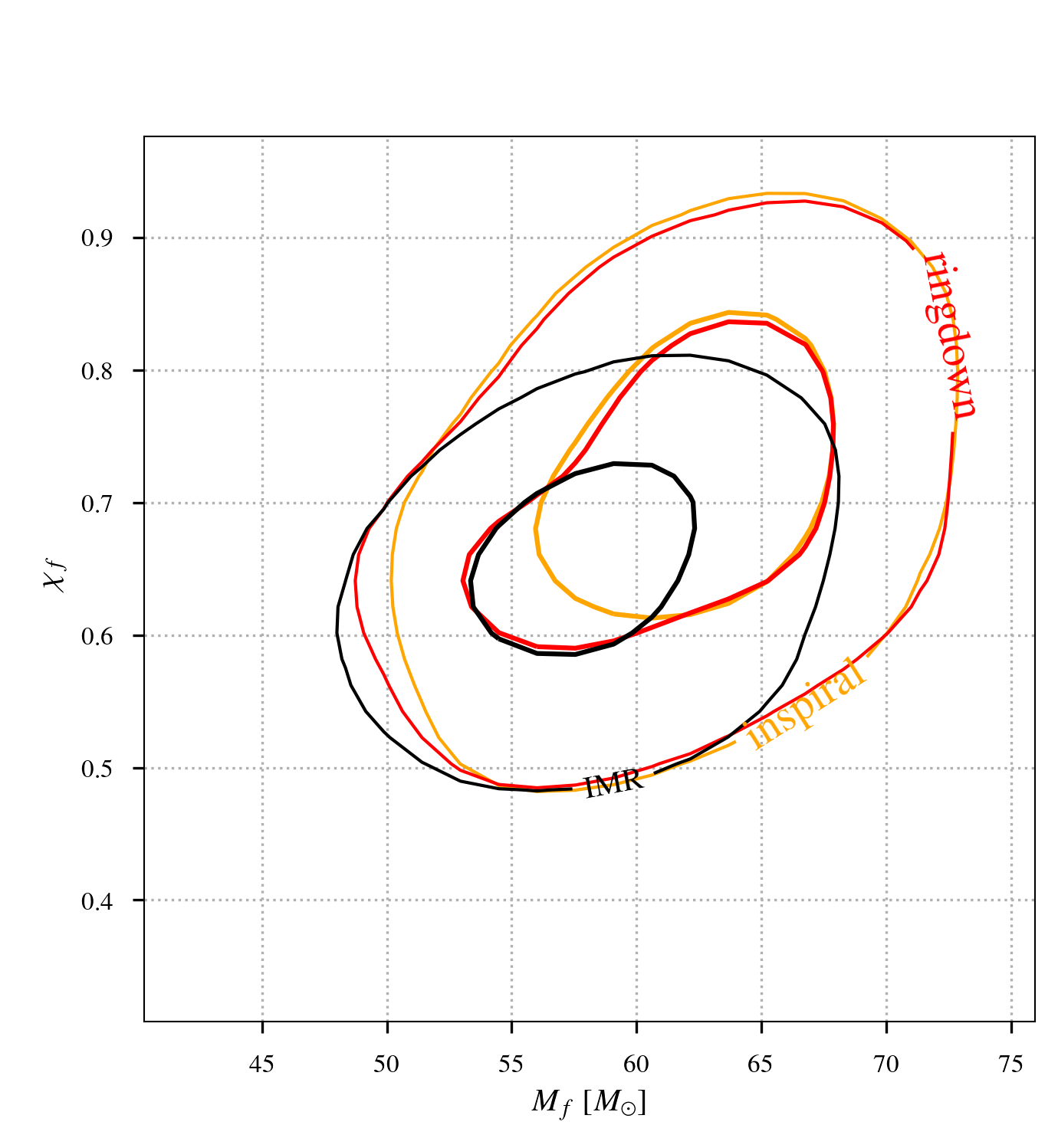}\includegraphics[width=7.5cm]{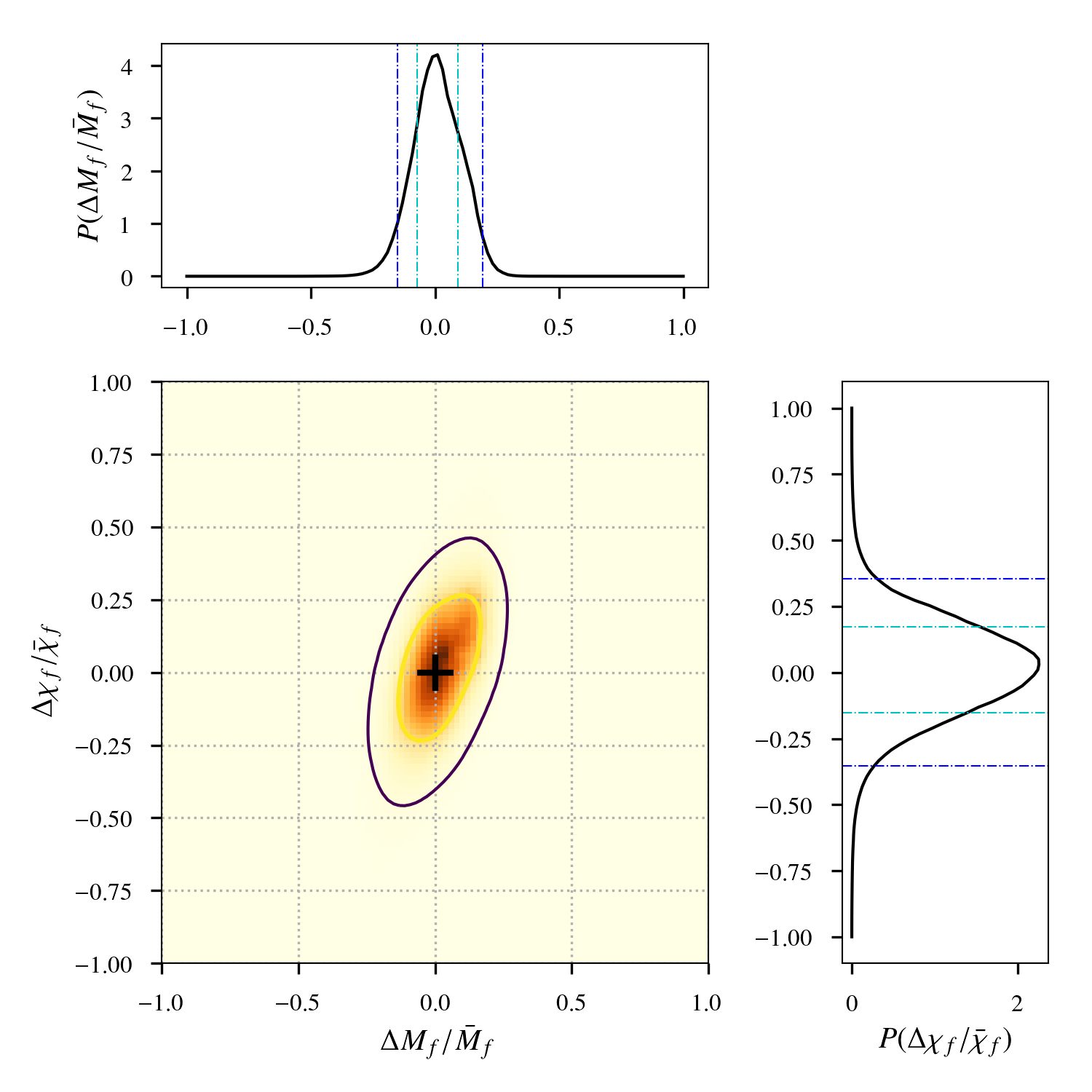}
\caption[IMR consistency test of GW170104 (pure NR)]{On the left, IMR overlap plot of GW170104 using all HMs up to $\ell = 4 $ with fitting over pure NR simulations assuming aligned spins; on the right, the joint probability distributions of $\Delta M_f/\bar M_f,\, \Delta \chi_f/\bar \chi_f$ does not show deviation from GR above 1.9\%. All the contours indicate the 68\% and the 95\% credible regions.}
\label{imrtgr_0104_nr}
\end{figure}

\subsubsection{GW170809}

\par{In this analysis we use a flat prior for the masses taking $M$ in the range between $45\,M_\odot$ and $105\,M_\odot$ and the spins components' prior are in agreement with the aligned spins assumption. Our model \texttt{NRSur7dq2} includes all HMs up to $\ell=4$ and the chosen cut-off frequency is 136 Hz, in accordance with the non-HMs analyses in Ref.~\cite{o2_tgr}.}

\subsubsection{GW170814}

\par{This observation leads to wide and bimodal posterior distributions for the post-inspiral analysis, as seen in Ref.~\cite{o2_tgr}. We use a flat prior for the masses components taking $M$ in the range between $10\,M_\odot$ and $80\,M_\odot$ and the spins components' prior are in agreement with the aligned spins assumption. The chosen cut-off frequency is 150 Hz, differently from the analysis in Ref.~\cite{o2_tgr} (where it is used $f_\text{cut}=161$ Hz). However, that is the results coming from our computations of ISCO frequency using the full IMR posterior samples. }

\par{We can see in Fig.~\ref{imrtgr_0814_surr} that the results do not deviate from GR, since the prediction is enclosed in the contour at 9.8\% credible region. However, the post-inspiral portion of data generates a multimodality with an additional peak at lower values of masses. The results from the inspiral and the post-inspiral signals are consistent with each other and they agree with the entire IMR's posterior distributions. We also note that the second peak is not able to exceed the 68\% confidence level of the distribution; however this fact is due to the restrictions on the spins which do not allow our posterior samples for the final quantities to reach that region of the parameters space; i.e. because of the choice of aligned spins prior which favors small BH spins.} 

\par{We repeat the measures using pure NR with an uniform distribution for the masses $m_1$ and $m_2$ in the region corresponding to $\eta\in [0.01,0.25]$ and $M \in [10\,M_\odot , 80\,M_\odot]$ assuming aligned spins in the range $a_{i,z}\in[-1,+1]$ for $i =1,2$. As it is shown in Fig.~\ref{imrtgr_0814_nr}, these results are almost identical, modulo sampling uncertainties in both calculations: a secondary peak still appear at lower masses' values and also increasing the number of iterations we are not able to remove it. In general, even if we get a bimodality in the posterior distribution for the post-inspiral signal which is inconsistent with the full-signal analysis, the GR quantiles in enclosed in the 90\% of confidence level and more advanced studies~\cite{o2_tgr} show that these results are reasonable with noise fluctuations.}

\subsubsection{GW170818}

\par{We use a flat prior for the masses taking $M$ in the range between $50\,M_\odot$ and $100\,M_\odot$ and the spins components' prior are in agreement with the aligned spins assumption. Our model \texttt{NRSur7dq2} includes all HMs up to $\ell=4$ and the chosen cut-off frequency is 128 Hz.}

\par{We note that for this event the constraint $q<2$ is a bit more stringent since a large number of inspiral's posterior samples coming from non-HMs analysis are in this area (about 50\%) and our posterior samples hit the boundary of the prior imposed by NR surrogate. However, the full-signal's posterior distribution shows that the maximum-posterior value respect this bound, and we are allowed to involve the NR surrogate. The results are in Fig.~\ref{imrtgr_0818_surr}.}

\begin{figure}[tb]
\centering
\includegraphics[width=7.5cm]{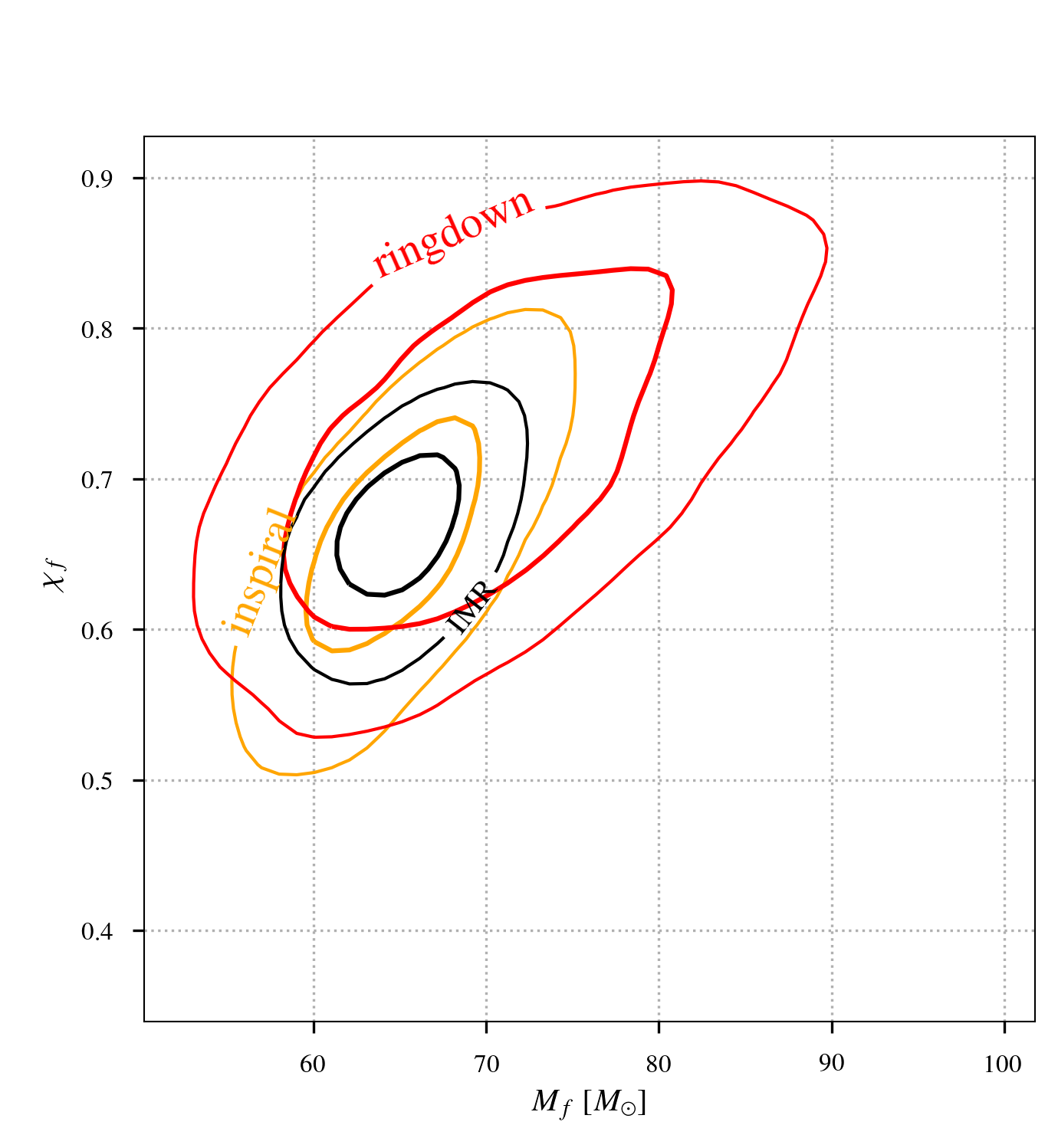}\includegraphics[width=7.5cm]{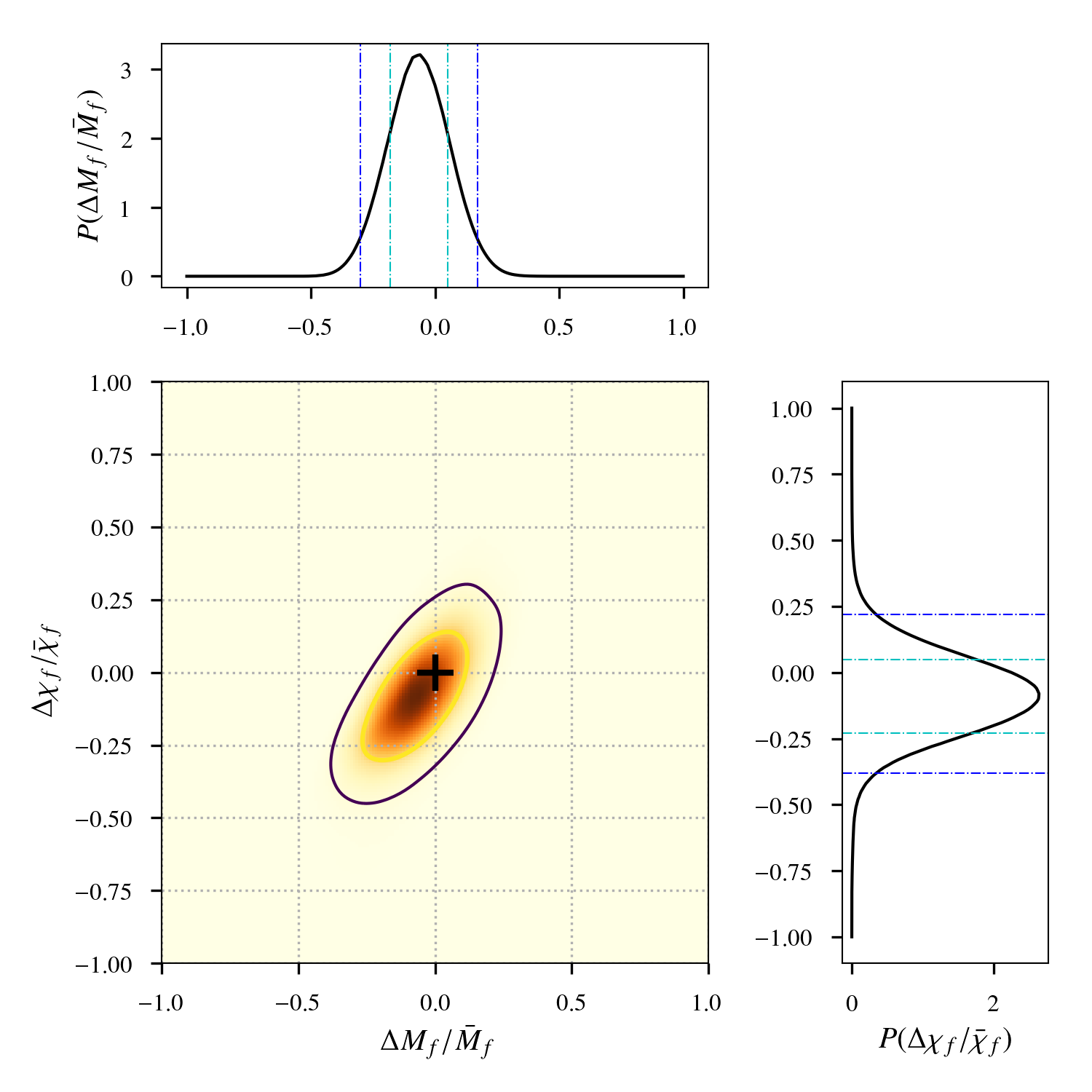}
\caption[IMR consistency test of GW170809 (NR surrogate)]{On the left, IMR overlap plot of GW170809 using all HMs up to $\ell = 4 $ with \texttt{NRSur7dq2} approximant assuming aligned spins; on the right, the joint probability distributions of $\Delta M_f/\bar M_f,\, \Delta \chi_f/\bar \chi_f$ does not show deviation from GR above 17.9\%. All the contours indicate the 68\% and the 95\% credible regions.}
\label{imrtgr_0809_surr}
\end{figure}

\begin{figure}[H]
\centering
\includegraphics[width=7.5cm]{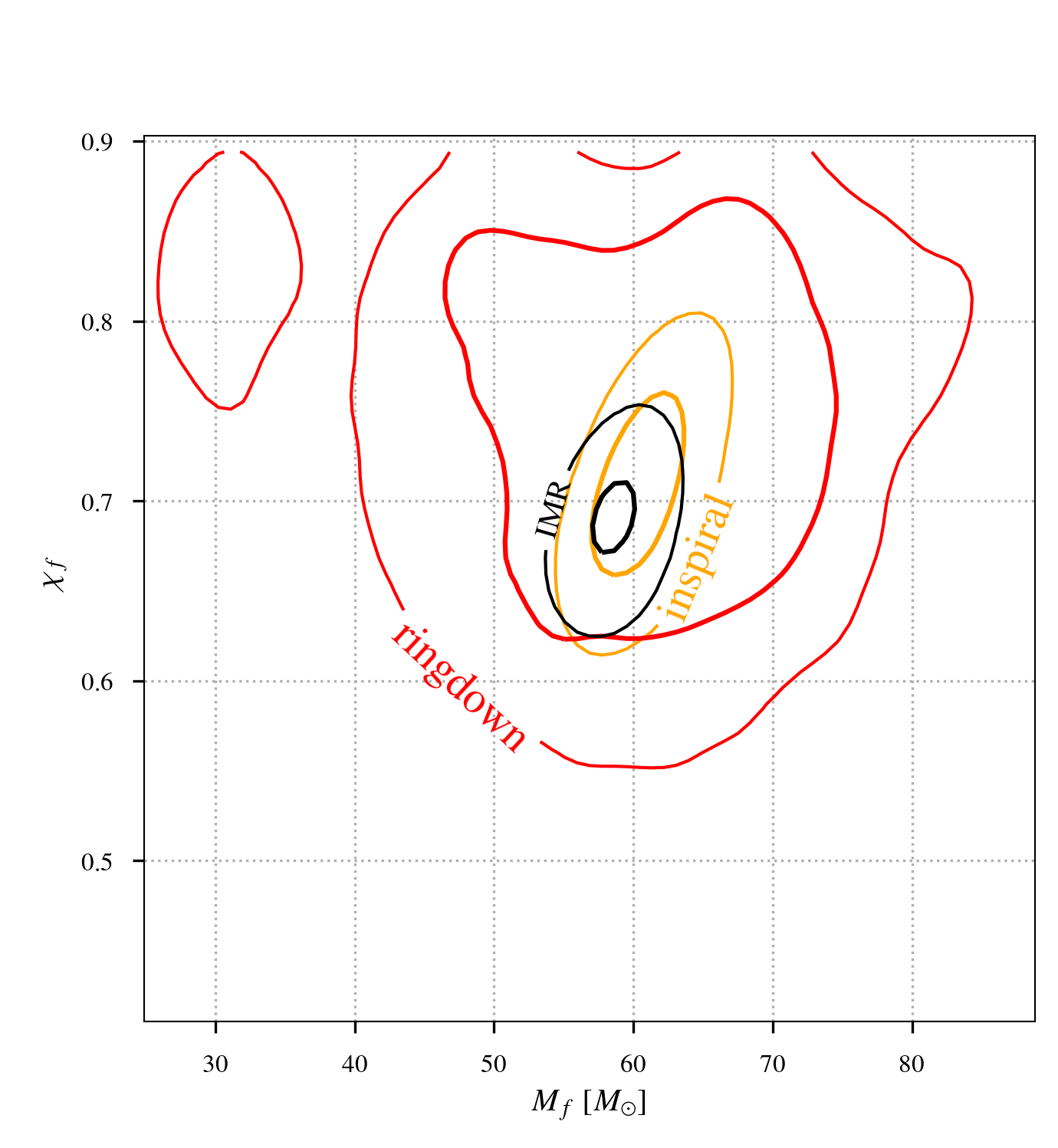}\includegraphics[width=7.5cm]{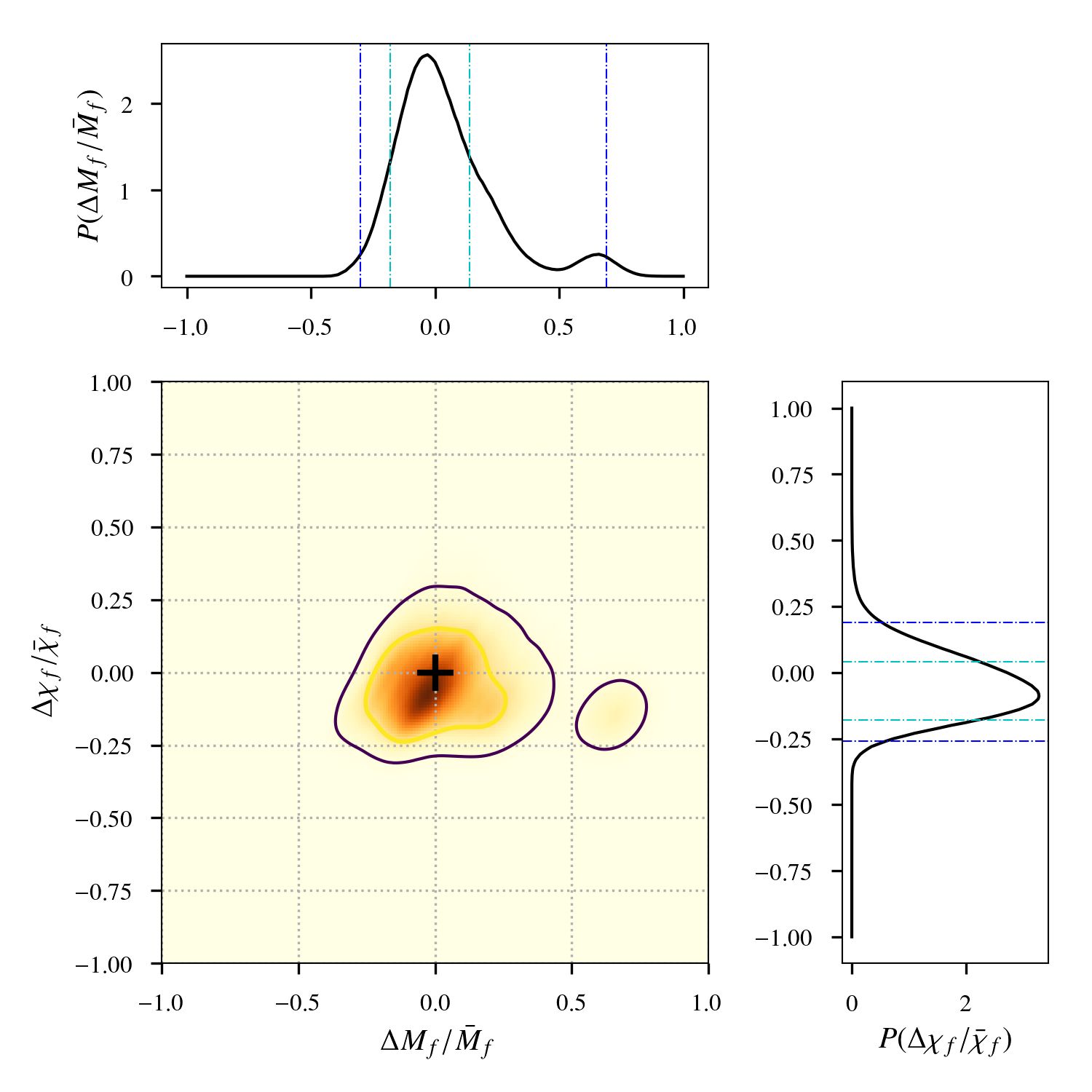}
\caption[IMR consistency test of GW170814 (NR surrogate)]{On the left, IMR overlap plot of GW170814 using all HMs up to $\ell = 4 $ with \texttt{NRSur7dq2} approximant assuming aligned spins; on the right, the joint probability distributions of $\Delta M_f/M_f,\, \Delta \chi_f/\bar \chi_f$ does not show deviation from GR above 9.8\%. All the contours indicate the 68\% and the 95\% credible regions.}
\label{imrtgr_0814_surr}
\end{figure}

\begin{figure}[H]
\centering
\includegraphics[width=7.5cm]{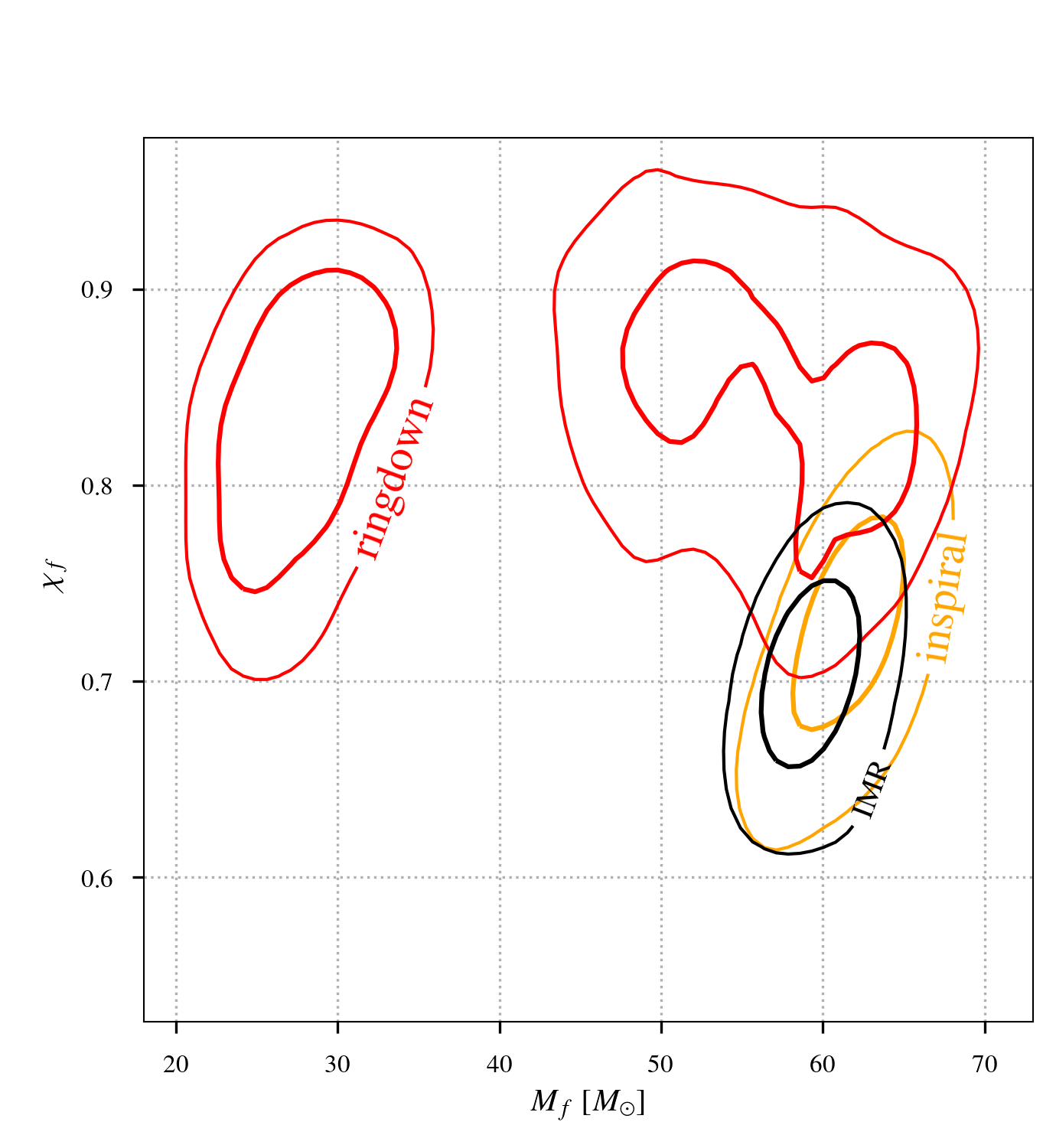}\includegraphics[width=7.5cm]{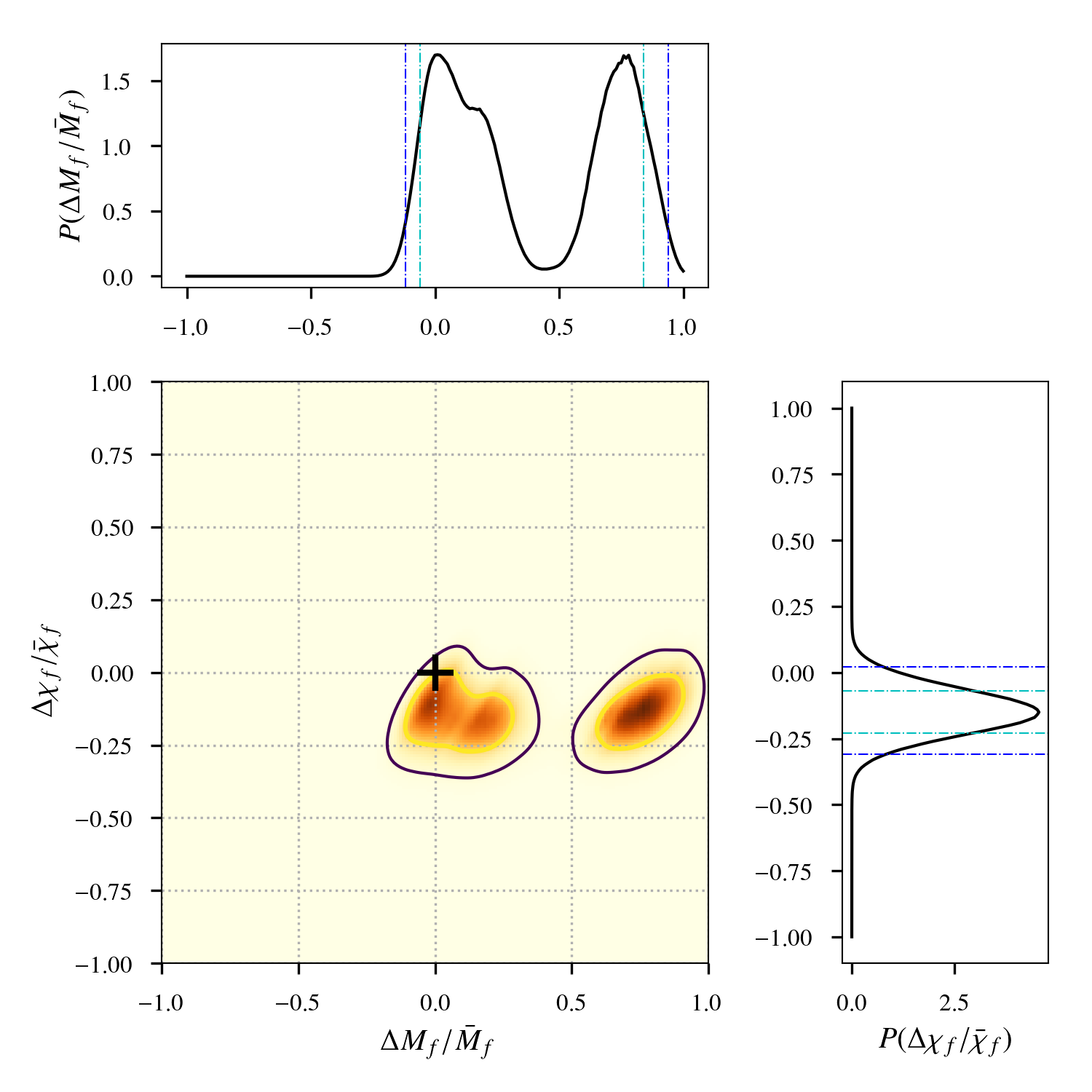}
\caption[IMR consistency test of GW170814 (pure NR)]{On the left, IMR overlap plot of GW170814 using all HMs up to $\ell = 4 $ fitting on pure NR simulations assuming aligned spins; on the right, the joint probability distributions of $\Delta M_f/\bar M_f,\, \Delta \chi_f/\bar \chi_f$ does not show deviation from GR above 72.1\%. All the contours indicate the 68\% and the 95\% credible regions.}
\label{imrtgr_0814_nr}
\end{figure}

\begin{figure}[H]
\centering
\includegraphics[width=7.5cm]{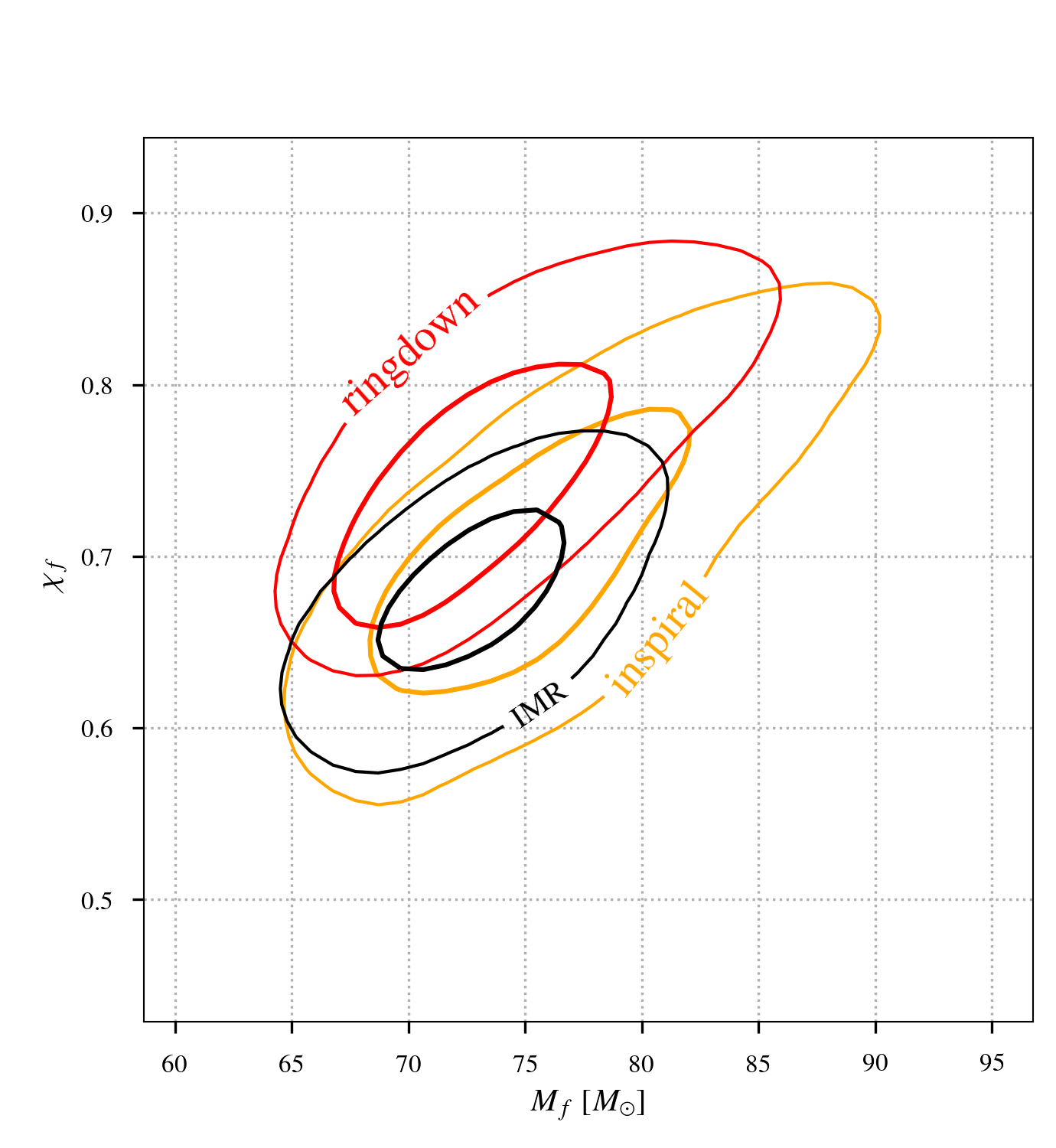}\includegraphics[width=7.5cm]{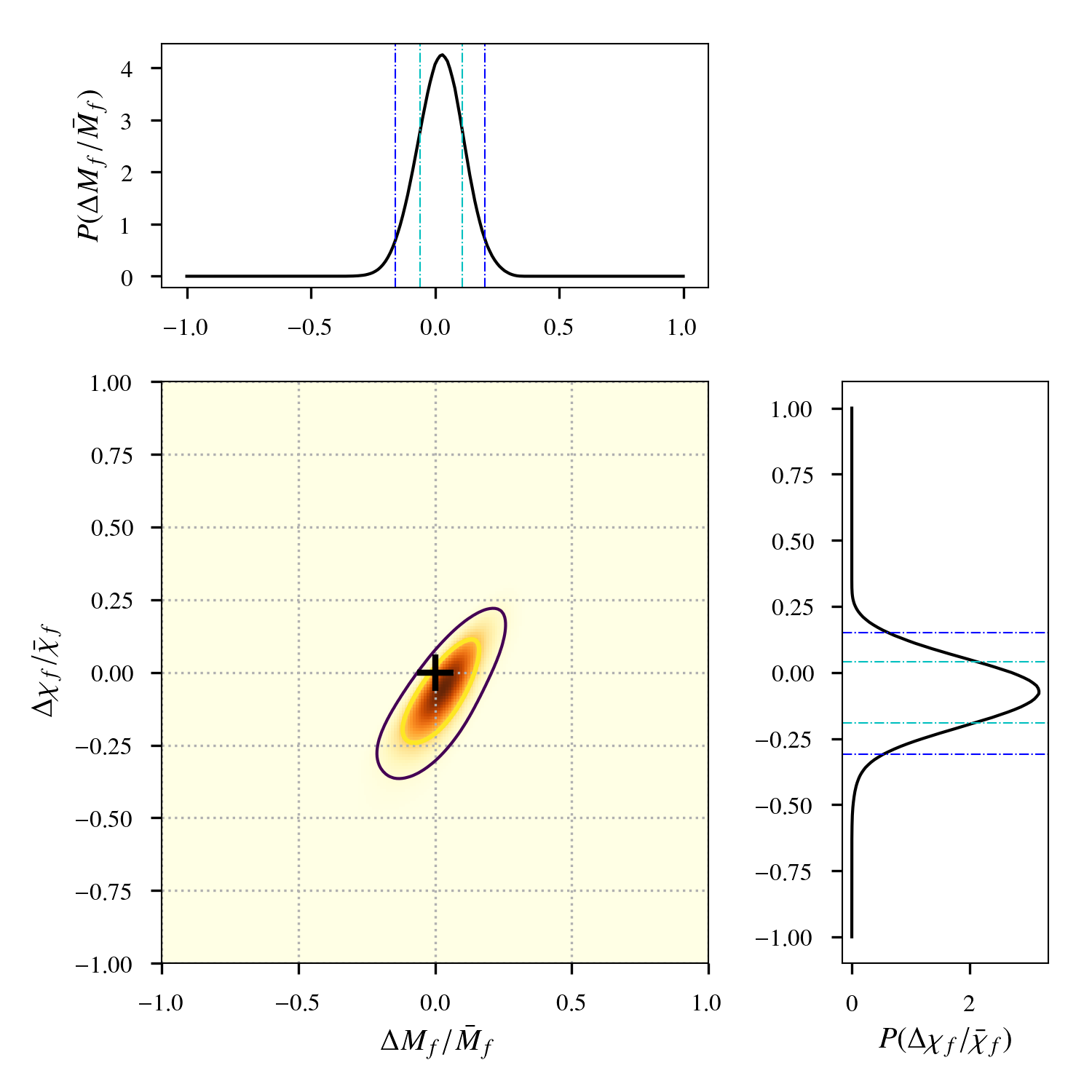}
\caption[IMR consistency test of GW170818 (NR surrogate)]{On the left, IMR overlap plot of GW170818 using all HMs up to $\ell = 4 $ with \texttt{NRSur7dq2} approximant assuming aligned spins; on the right, the joint probability distributions of $\Delta M_f/\bar M_f,\, \Delta \chi_f/\bar \chi_f$ does not show deviation from GR above 52.7\%. All the contours indicate the 68\% and the 95\% credible regions.}
\label{imrtgr_0818_surr}
\end{figure}

\subsubsection{GW170823}

\par{Also GW170823 leads to multimodalities in the posterior distributions but, unlike GW170814, in this case they come from the inspiral's portion of data. This fact could come from the fact that we observed few inspiralling orbits and this leads to large parameters uncertainties. We use a flat prior for the masses $m_1$ and $m_1$ imposing $M$ in the range between $50\,M_\odot$ and $220\,M_\odot$ and the spins components' prior are in agreement with the aligned spins assumption. Our model \texttt{NRSur7dq2} includes all HMs up to $\ell=4$ and the chosen cut-off frequency is 102 Hz.}

\par{In Fig.~\ref{imrtgr_0823_surr} we can see that the results involving HMs shows the degeneracy at higher mass in the inspiral's posterior distribution, however the additional peak is not the one with higher probability. Furthermore, we involve pure NR simulations using an uniform prior distribution for the masses components $m_1$ and $m_2$ in the region corresponding to $\eta\in [0.01,0.25]$ and $M \in [50\,M_\odot , 220\,M_\odot]$, we assume aligned spins in the range $a_{i,z}\in[-1,+1]$ for $ i =1,2$. Also in this case we use all HMs with $\ell\le 4$. In Fig.~\ref{imrtgr_0823_nr} we can see that the secondary peak reach almost the 68\% credible region and contains a relatively small portion of the entire posterior volume. }

\begin{figure}[tb]
\centering
\includegraphics[width=7.5cm]{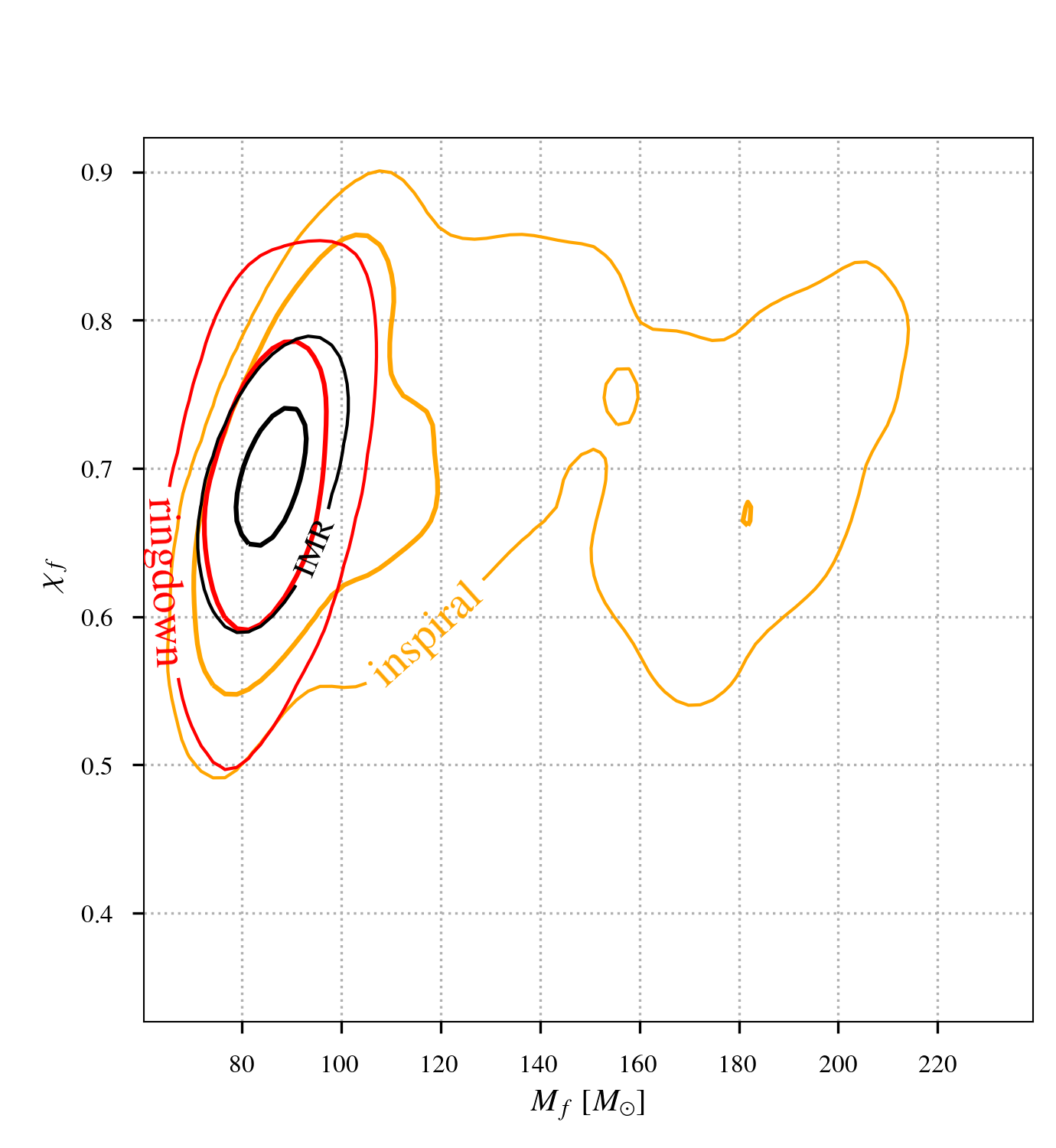}\includegraphics[width=7.5cm]{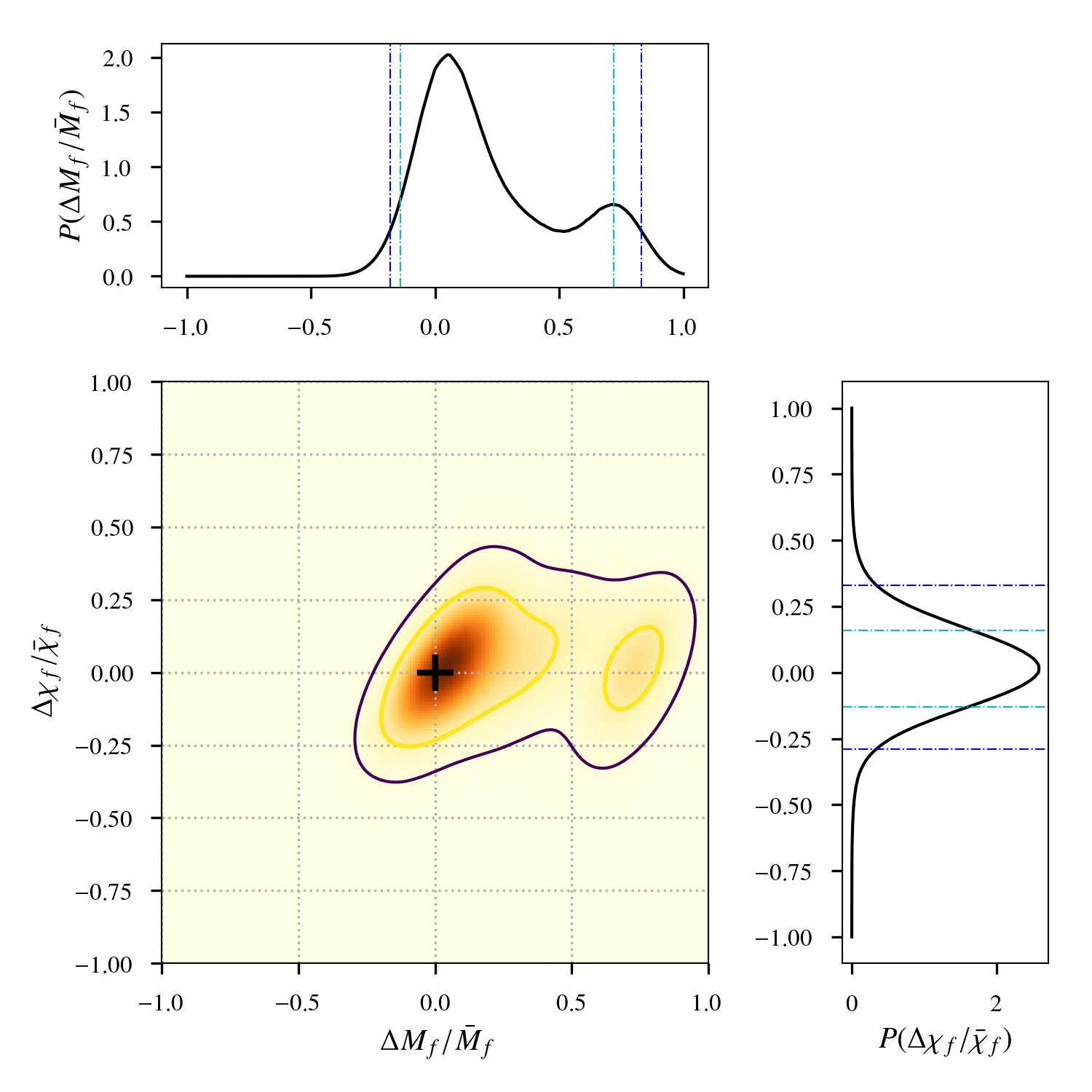}
\caption[IMR consistency test of GW170823 (NR surrogate)]{On the left, IMR overlap plot of GW170823 using all HMs up to $\ell = 4 $ with \texttt{NRSur7dq2} approximant assuming aligned spins; on the right, the joint probability distributions of $\Delta M_f/\bar M_f,\, \Delta \chi_f/\bar \chi_f$ does not show deviation from GR above 2.9\%. All the contours indicate the 68\% and the 95\% credible regions.}
\label{imrtgr_0823_surr}
\end{figure}

\begin{figure}[H]
\centering
\includegraphics[width=7.5cm]{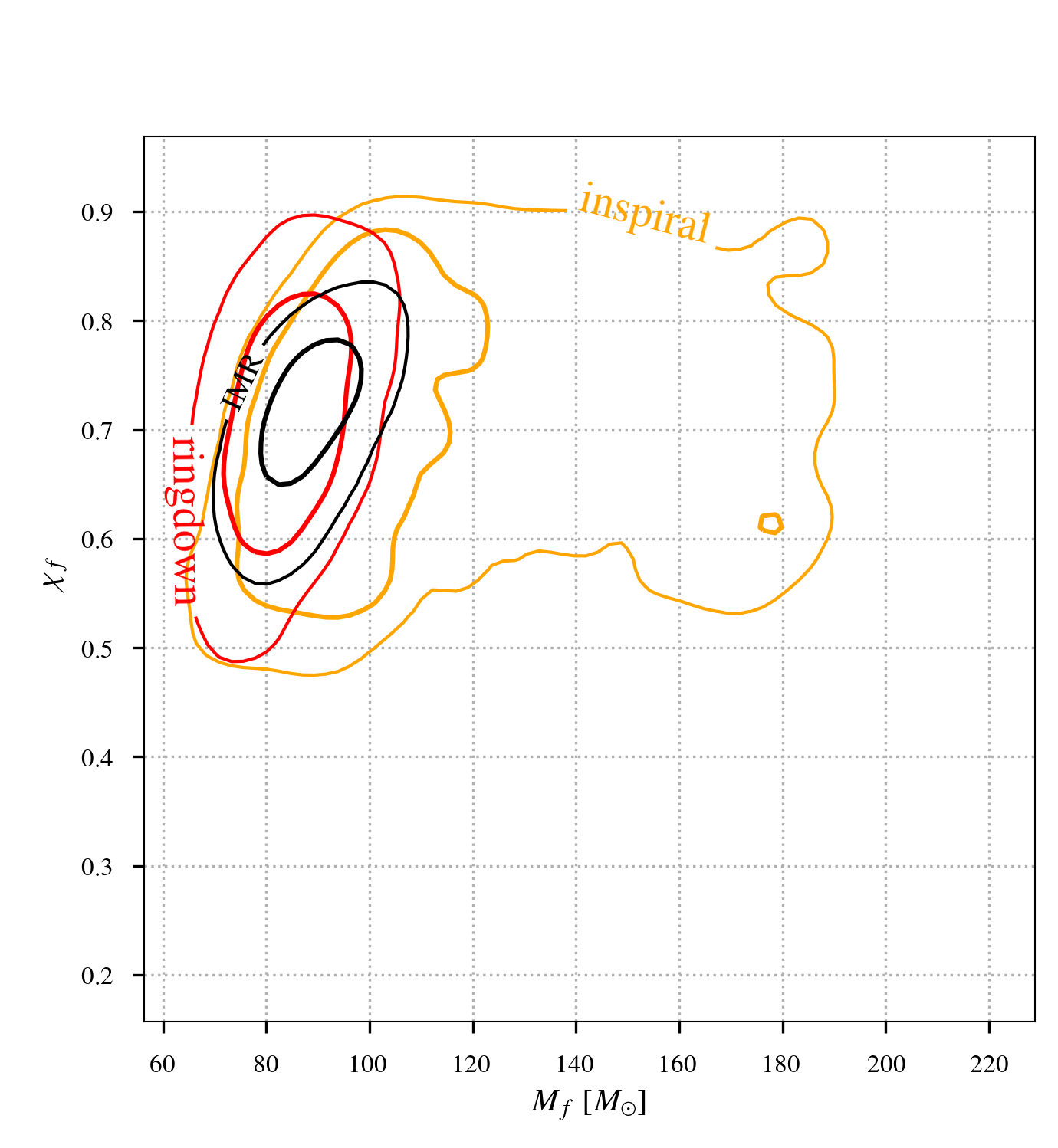}\includegraphics[width=7.5cm]{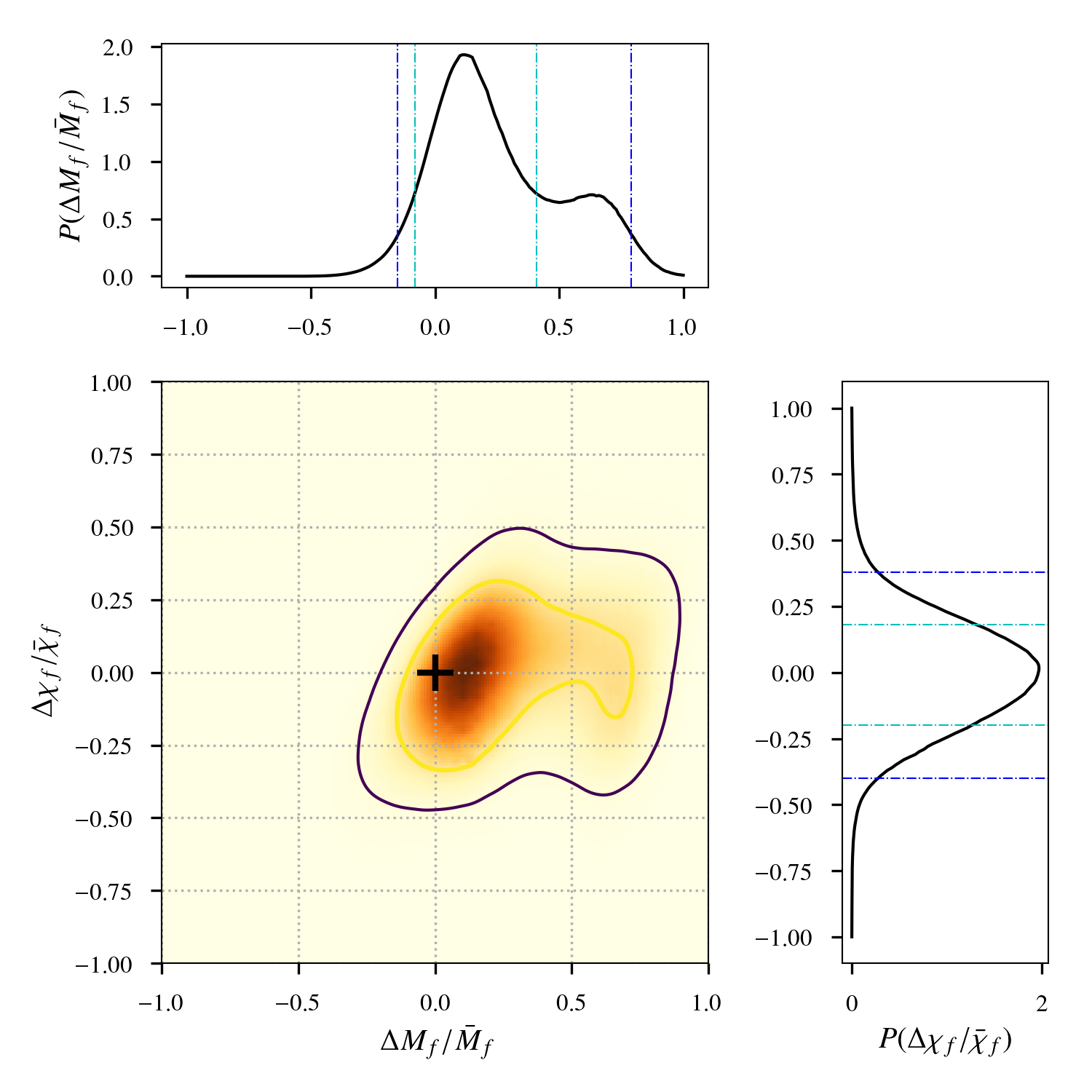}
\caption[IMR consistency test of GW170823 (pure NR)]{On the left, IMR overlap plot of GW170823 using all HMs up to $\ell = 4 $ fitting on pure NR simulations assuming aligned spins; on the right, the joint probability distributions of $\Delta M_f/\bar M_f,\, \Delta \chi_f/\bar \chi_f$ does not show deviation from GR above 22.3\%. All the contours indicate the 68\% and the 95\% credible regions.}
\label{imrtgr_0823_nr}
\end{figure}

\section{Conclusions}

\par{We note that the IMR tests in Ref.~\cite{o2_tgr} performed with the approximant \texttt{IMRPhenomPv2}~\cite{imrphenom, imrphenom2, imrphenom3} and \texttt{SEOBNRv4\_ROM}~\cite{seobnrv4} show the same degeneracies coming from the posterior distributions of $M_f^\text{MR}$ for GW170814 and from $M_f^\text{I}$ for GW170823. However this contributions do not show the presence of coherent signals inside the data and they are reasonable with noise fluctuations. }

\par{Then we use the defined quantities $\Delta M_f/\bar M_f$ and $\Delta \chi_f/\bar \chi_f$ to join together the predictions coming from different events. So, we take the results coming from IMR tests of all O2 events where we used the NR surrogate, obtaining a single distribution which tells us the combined prediction of these events. The join posterior distribution does not show deviations from GR above the 39.3\% confidence level, and we can see in Fig.~\ref{join_distribution} the the $+$ symbol is largely enclosed in the contour at 90\% credible region. This results is totally in accordance with the expected results we are able to infer that the inclusion of HMs in the IMR consistency test does not allow to deviation from the GR prediction. Then we combined the results from the analyses which involve pure NR. The posterior distributions are shown in Fig.~\ref{join_distribution_nr} and the join distribution does not show deviation above the 55.1\% confidence level.}

\begin{figure}[tb]
\centering
\includegraphics[width=9cm]{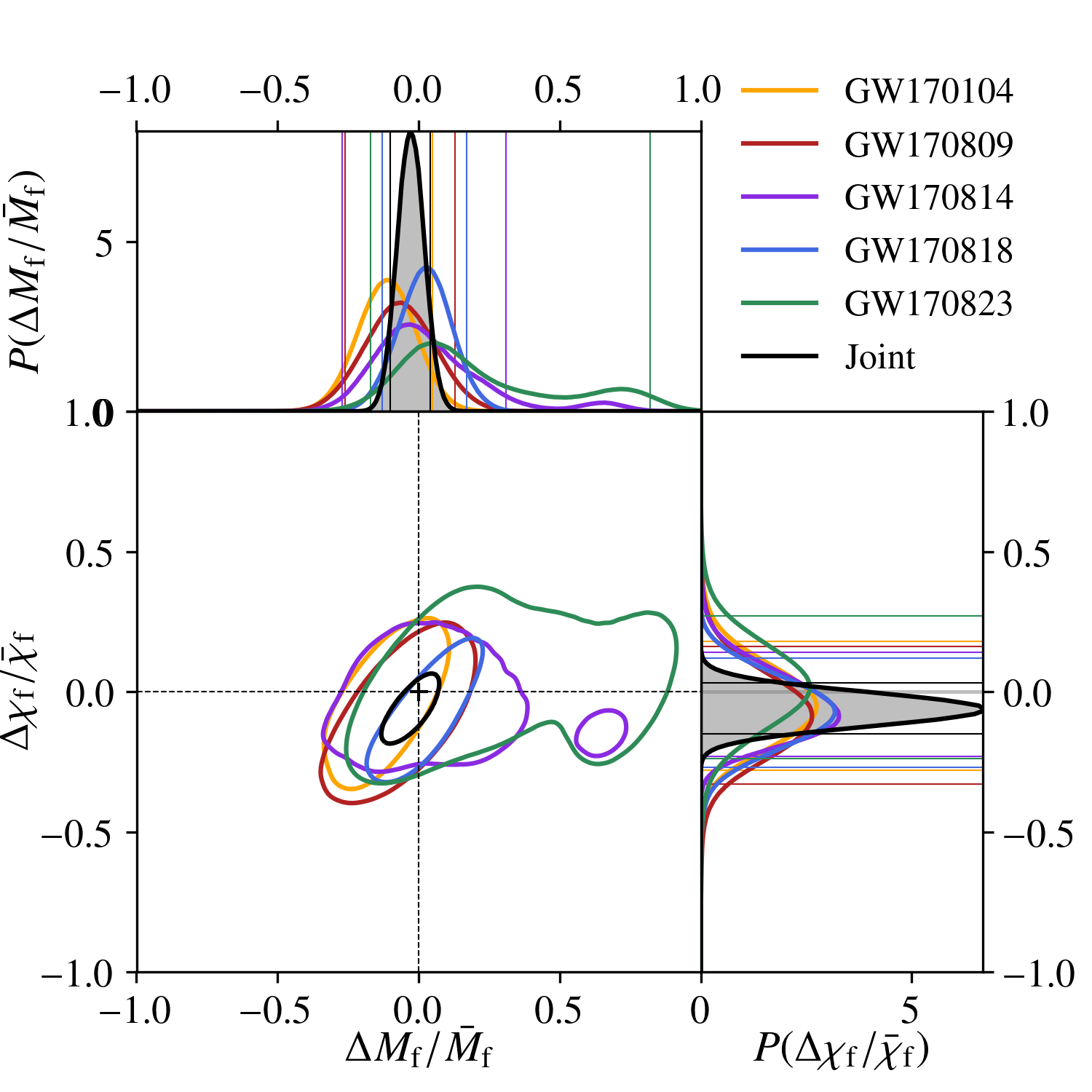}
\caption[Combined posterior distribution for O2 events with NR surrogate]{The 90\% credible region of the join posterior distribution and the relative marginalization from the analyses involving \texttt{NRSur7dq2}. These combined results do not show deviation above the 39.3\% credible region. The symbol $+$ indicates GR prediction.}
\label{join_distribution}
\end{figure}

\begin{figure}[tb]
\centering
\includegraphics[width=9cm]{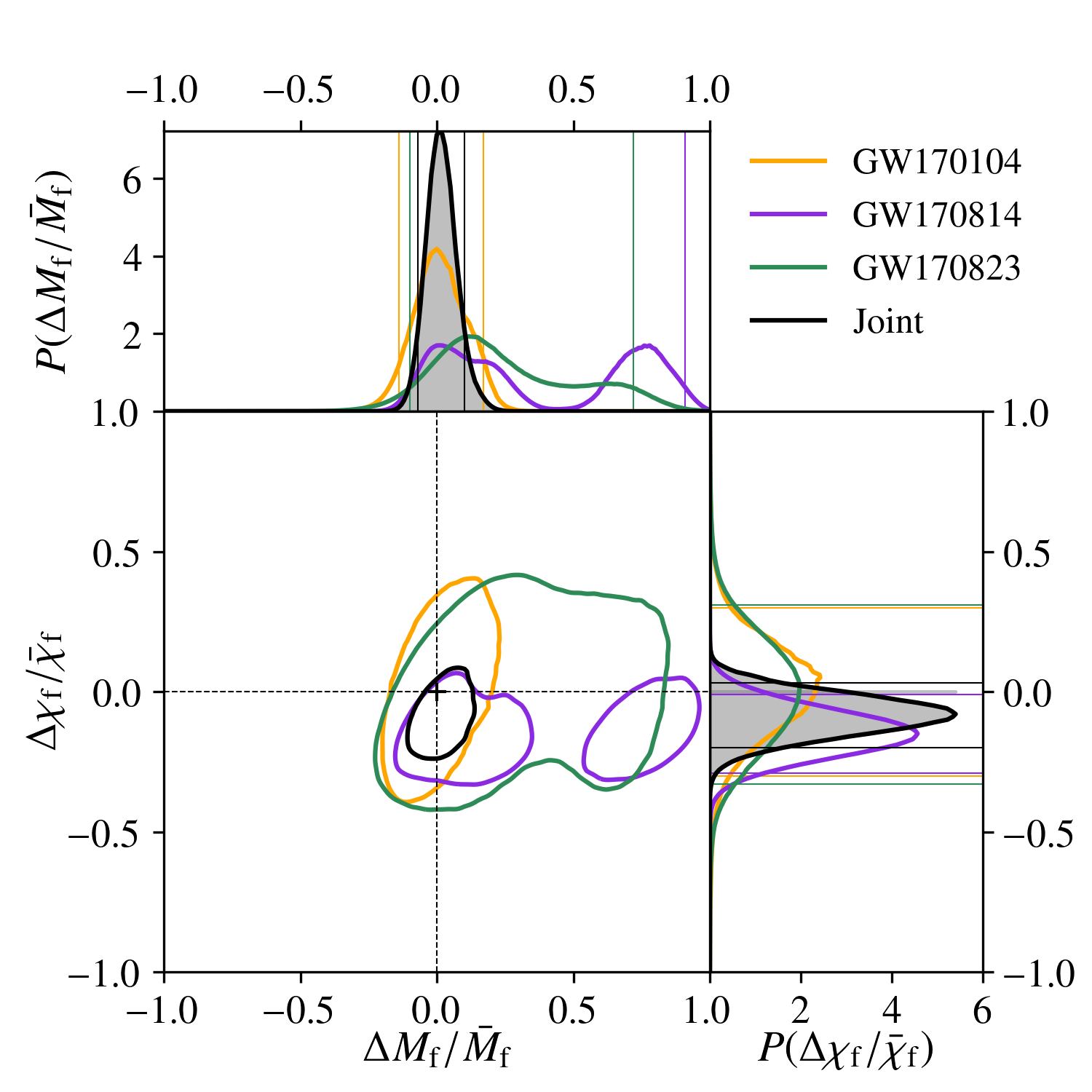}
\caption[Combined posterior distribution for O2 events with pure NR]{The 90\% credible region of the join posterior distribution and the relative marginalization from the analyses involving pure NR. These combined results do not show deviation above the 55.1\% credible region. The symbol $+$ indicates GR prediction.}
\label{join_distribution_nr}
\end{figure}

\section*{Acknowledgment}

\par{The authors are grateful for computational resources provided by the LIGO Laboratory and supported by National Science Foundation Grants PHY-0757058 and PHY-0823459. We also thank Dr. Abhirup Ghosh, Dr. Rahul Kashyap and Dr. Nathan K. Johnson-McDaniel for their useful comments. MB would like to thank the INFN that granted him the scholarship for the exchange program at RIT and he acknowledges support by the EU H2020 under ERC Starting Grant, no. BinGraSp-714626.}

$\quad$\\
E-mail: \href{mailto:matteo.breschi@ligo.org}{matteo.breschi@ligo.org}

 \end{document}